\documentclass[12pt]{iopart}
\usepackage{iopams}  
\expandafter\let\csname equation*\endcsname\relax

\expandafter\let\csname endequation*\endcsname\relax

\usepackage{amsmath}

\usepackage{stmaryrd} 
\usepackage{mathcomp} 
\usepackage{hyperref} 

 \usepackage{bm}
 \usepackage{graphicx}
 \usepackage{latexsym}
 \usepackage{enumerate}

\newcommand{\grad}{ \bm{\nabla}} 
\newcommand{\divv}{ \bm{\nabla\cdot}} 
\newcommand{\curl}{ \bm{\nabla\times}} 
\newcommand{\ijot}{ \mathrm{i} } 
\renewcommand{\e}{ \mathrm{e} } 
\renewcommand{\d}{ \mathrm{d} } 
\newcommand{\esub}[1]{\bm{e}_{#1}} 
\newcommand{\dotv}{  \bm{\cdot} } 
\newcommand{\cross}{  \bm{\times} } 

\newcommand{\sgn} {\mathrm{sgn}\,}
\newcommand{\sech} {\mathrm{sech}}
\renewcommand{\Re} {\mathrm{Re}\,}
\renewcommand{\Im} {\mathrm{Im}\,}

\newcommand{\const}{{\mathrm{const}}}

\newcommand{\vrm}[1]{{\boldsymbol{#1}}} 

\newcommand{\jump}[1]{\left\llbracket  #1 \right\rrbracket} 

\newcommand{\slt}{{\scriptscriptstyle <}}
\newcommand{\sgt}{{\scriptscriptstyle >}}
\newcommand{\slgt}{{\scriptscriptstyle \lessgtr}} 
\newcommand{\sglt}{{\scriptscriptstyle \gtrless}} 
\newcommand{\signgt}{\sgn\!({\scriptstyle >})}
\newcommand{\signlt}{\sgn\!({\scriptstyle <})}
\newcommand{\signglt}{\sgn\!({\scriptstyle\gtrless})}
\newcommand{\signlgt}{\sgn\!({\scriptstyle\lessgtr})}

\newcommand{\muSI}{\tcmu_0}
\newcommand{\iotabar}{\mbox{$\,\iota\!\!$-}}

\begin{document}

\title[MRxMHD Spectrum]{Spectrum of multi-region-relaxed magnetohydrodynamic modes in topologically toroidal geometry}

\author{R.~L. Dewar, L.~H. Tuen, M.~J. Hole}

\address{Centre for Plasmas and Fluids, Research School of Physics \& Engineering, The Australian National University, Canberra, ACT 2601, Australia}
\ead{robert.dewar@anu.edu.au}
\vspace{10pt}
\begin{indented}
\item[Version 2.0: ] 13 November 2016  
\end{indented}

\begin{abstract}
A general formulation of the problem of calculating the spectrum of stable and unstable eigenmodes of linearized perturbations about a magnetically confined toroidal plasma is presented. The analysis is based on a new hydromagnetic dynamical model, Multi-region Relaxed Magnetohydrodynamics (MRxMHD), which models the plasma-magnetic field system as consisting of multiple regions, containing compressible Euler fluid and Taylor-relaxed magnetic field, separated by interfaces in the form of flexible ideal-MHD current sheets. This is illustrated using a first-principles analysis of a two-region slab geometry, with periodic boundary conditions to model the outer regions of typical tokamak or reversed-field pinch plasmas. The lowest and second-lowest eigenvalues in plasmas unstable to tearing and kink-tearing modes are calculated.  Very near marginal stability the lowest mode obtained using the incompressible approximation to the kinetic energy normalization of the present study is shown to correspond to the eigenvalues found in previous studies where all mass was artificially loaded onto the interfaces. 
\end{abstract}

%
%
\submitto{\PPCF}
%
%
%

\section{Introduction}\label{sec:intro}

Despite the complex many-particle nature of plasmas at short scales, the use of fluid models has been remarkably successful in understanding the larger-scale collective behaviour of plasmas and such models are still routinely used in modelling magnetically confined fusion plasmas. The simplest such models \cite{Freidberg_82} treat the plasma as a single fluid and apply on length scales much longer than a typical ion gyroradius and time scales longer than a typical inverse ion cyclotron frequency, and also sufficiently long for Maxwell's displacement current to be negligible. 


We may term all such theories magnetohydrodynamics (MHD), but in this paper we consider only ideal and \emph{relaxed} MHD, the latter being obtained from ideal MHD by removing most of its microscopic constraints. All  MHD models can be described in the \emph{Eulerian picture}, which describes the fluid dynamics in terms of the space- and time-dependent \emph{fields} --- mass density $\rho(\vrm{x},t)$, isotropic pressure $p(\vrm{x},t)$, and magnetic field $\vrm{B}(\vrm{x},t)$. Ideal MHD can also be described in the \emph{Lagrangian picture}, which regards the fluid as consisting of moving \emph{fluid elements} of volume $\d V^t$, position $\vrm{r}^t(\vrm{r}_0)$ and mass $\rho(\vrm{r}^t,t)\d V^t$, with $\rho$, $p$ and $\vrm{B}$ holonomically (microscopically) constrained to be advected with the fluid elements. Relaxed MHD can be described as a hybrid theory where the Lagrangian picture can be used to describe $\rho$ as in ideal MHD, but the fields $p$ and $\vrm{B}$ must be treated in the Eulerian picture as they are subjected only to non-holonomic, macroscopic constraints.\footnote{Another hybrid  case is resistive MHD, where $p$ may be holonomically advected but $\vrm{B}$ diffuses as a free field, at least in a thin resonant reconnection layer.}

Although the concept of plasma self-organization through a relaxation process describable by a variational (energy-minimization) principle is an old one (see e.g. the review by Taylor \cite{Taylor_86}), the generalization to a fully fledged fluid theory, \emph{Multi-region Relaxed Magnetohydrodynamics} (MRxMHD) has only recently been enunciated \cite{Dewar_Yoshida_Bhattacharjee_Hudson_15,Dewar_Hudson_Bhattacharjee_Yoshida_16}. 
Unlike previous, quasi-static, multi-region generalizations of Taylor relaxation \cite{Hudson_etal_12b}, this new formulation is a \emph{fully dynamical}, time-dependent field theory whose self-consistency is ensured by deriving it from an action principle rather than an energy principle.
	
The deficiencies of ideal MHD for describing typical fusion plasmas \cite{Freidberg_82} arise from its assumptions of (a) zero thermal conductivity, which implies ``frozen-in'' entropy (i.e. an adiabatic equation of state applying in each fluid element); and of (b) infinite electrical conductivity, which implies \cite{Newcomb_58} frozen-in magnetic field. Assumption (a) is clearly inapplicable in high-temperature plasmas as electron mean-free-paths along magnetic field lines are very long, making parallel heat conduction large. On the other hand, electrical conductivity is indeed large in fusion-relevant plasmas so assumption (b) is at first sight a reasonable approximation.

However resistivity and other non-ideal effects are enhanced in regions with short scale lengths, such as current sheets and resonances, giving rise to changes in magnetic-field-line topology (reconnection) on mesoscopic timescales, such as the growth of magnetic islands through tearing instabilities arising at resonant magnetic surfaces. Such topological changes are forbidden by the frozen-in magnetic flux property of ideal MHD, motivating the search for a simple fluid model for fusion plasmas that is more appropriate physically than ideal MHD.

The fundamental problem with ideal MHD is that conserving entropy and magnetic flux separately in an uncountable infinity of fluid elements makes it physically over-constrained, which is resolved in MRxMHD by using only a \emph{subset} of ideal MHD's constraints. This subset consists of entropy and magnetic helicity constraints within an arbitrary number of finite sub-volumes of the plasma, plus ideal-MHD boundary constraints on the infinity of surface elements making up the interfaces between these subregions. The parsimonious choice made in \cite{Dewar_Yoshida_Bhattacharjee_Hudson_15}, to use only magnetic helicities rather than magnetic helicities plus fluid-magnetic cross helicities, as volume constraints decouples the magnetic field from the fluid within the subregions, giving a very simple generalization of Taylor relaxation in which the plasma behaves as an Euler fluid within each subregion. 

A principal motivation of the development of MRxMHD has been the need for a better MHD framework than ideal MHD for numerical calculations of equilibria in fully three-dimensional magnetic containment devices such as stellarators, and tokamaks with resonant magnetic perturbations, where smoothly nested flux surfaces cannot be assumed. The theoretical development and physical application of this important static application is already well developed \cite{Hudson_etal_12b}. 

It is anticipated that the new dynamical formulation \cite{Dewar_Yoshida_Bhattacharjee_Hudson_15} will likewise provide a better framework than ideal MHD for efficient numerical calculations of stability and mode structure in realistic geometries. However, as the decoupling of fluid flow and magnetic field except at discrete interfaces seems at first sight a dramatic oversimplification, confidence that this novel formulation is physically reasonable needs to be built up through the analysis of simple test cases with a few interfaces, where fundamental physical effects can be isolated and analyzed in detail. Then it will need to be shown that, in the limit of an unbounded number of interfaces, MRxMHD approaches a physically applicable continuum theory that is equivalent or superior to ideal MHD (cf. \cite{Dennis_Hudson_Dewar_Hole_13}). This paper represents a first step in this program of research. 

\begin{figure}[htbp]
   \centering
		\includegraphics[width = 0.7\textwidth]{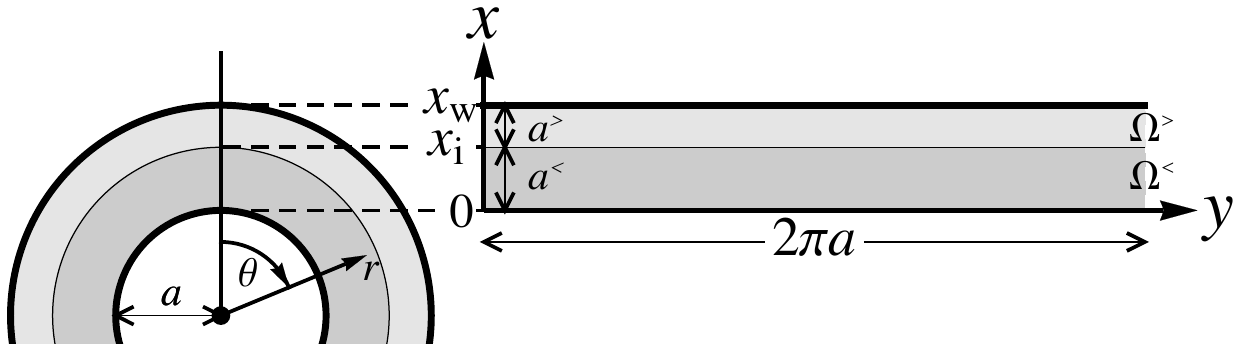} 
\caption{Mapping the outer two regions of a 3-region cylindrical plasma equilibrium to our 2-region slab model. The $z$-axis is into the page. The slab plasma inherits the annular topology of the outer regions of the cylinder through application of periodic boundary conditions with periodicity length $L_{\rm pol} = 2\pi a$.}
\label{fig:CylSlab}
\end{figure}

It is the aim of the present paper to formulate a basic framework for calculating normal modes of linearized perturbations about general plasma equilibria using dynamical MRxMHD theory \cite{Dewar_Yoshida_Bhattacharjee_Hudson_15}. We also illustrate the application of this framework in a simple geometry to build physical insight into novel features of the MRxMHD spectrum. These insights will aid in the interpretation of the results of further studies in more realistic geometries. 

For maximum simplicity we consider a simple slab geometry, in which the equilibrium quantities are independent of periodic $y$ and $z$ Cartesian coordinates, the system being bounded by a perfectly conducting planar wall at $x = x_{\rm w}$ and, at $x = 0$, a rigid, perfectly conducting interface between the slab plasma and a notional cylindrical core plasma of radius $a$. 

This is depicted in Fig.~\ref{fig:CylSlab}, which shows how the slab model may be associated with the plasma (and possibly vacuum) between the inner core and the outer confining wall of a large-aspect-ratio toroidal plasma, approximated by a cylinder with periodic boundary conditions in the toroidal, $z$-direction. The slab model ``straightens out'' the poloidal direction, but keeps torus topology by applying periodic boundary conditions.
This toroidal-confinement analogy is not meant to imply that the slab model is adequate for experimental comparisons, but rather to allow us to use terminology and notation familiar in the toroidal confinement field, and to suggest orders of magnitude for parameters relevant to tokamak and reversed-field-pinch illustrative cases. 

We further simplify by assuming the plasma slab to comprise only two MRxMHD regions, a Taylor-relaxed inner plasma region $\Omega^\slt$, between $x = 0$ and an interface at $x = x_{\rm i} < x_{\rm w}$, and an outer region $\Omega^\sgt$ between $x = x_{\rm i}$ and $x_{\rm w}$. (We denote parameters belonging to the inner and outer regions by superscript ${}^\slt$ and ${}^\sgt$, respectively.) The outer region can either be a vacuum or another Taylor-relaxed region. The perturbed eigenfunctions (similar to the well-known ABC solutions \cite[see e.g.]{Dombre_etal_1986}) are represented in an elementary fashion as sums of complex plane waves obeying separate local dispersion relations in the plasma(s) or vacuum. The perturbations in the inner and outer regions are coupled via a surface wave on the interface between the regions.


In the following we give a brief summary of the new MRxMHD equations in general, Sec.~\ref{sec:MRxMHD}, and linearized, Sec.~\ref{sec:linGen},  form.  We then, Sec.~\ref{sec:slabgeom}, motivate and develop the cylindrical core + slab model and present illustrative tokamak-like and pinch-like cases. Wave perturbations of such equilibria are developed in Sec.~\ref{sec:slabeqwaves}, first calculating the modulations of the entropy and magnetic helicity Lagrange multipliers by modulating the interface position. We then treat plane waves (including evanescent waves) within the plasma and vacuum with wave vectors in general directions compatible with the periodic boundary conditions. 

These waves are superposed to find acoustic and magnetic standing waves in Sec.~\ref{sec:standing}, which are combined into the general eigenvalue problem in Sec.~\ref{sec:eigen}. The general spectrum in the case of a vacuum between $x = x_{\rm i}$ and $x = x_{\rm w}$ is examined graphically, computationally, and analytically in Sec.~\ref{sec:vac}. Tearing and kink-tearing modes are discussed in Sec.~\ref{sec:Tearing} and conclusions and ideas for future work are given in Sec.~\ref{sec:Concl}. An online supplementary version \cite{SuppCitePPCF} provides extra discussion and detail.

\section{The dynamical MRxMHD model}\label{sec:MRxMHD}
In \cite{Dewar_Yoshida_Bhattacharjee_Hudson_15} the equations for MRXMHD were derived as Euler--Lagrange equations from the Lagrangian
\begin{equation}\label{eq:LRx}
	L = \sum_i L_i - \int_{\Omega_{\rm v}} \frac{\vrm{B}\dotv\vrm{B}}{2\muSI} \,\d V
\;,
\end{equation}
where the volume integration $\int\! \d V$ is over a vacuum region $\Omega_{\rm v}$, with $\vrm{B}$ denoting magnetic field and $\muSI$ the permeability of free space.  The sum $\sum_i$ is over Lagrangians $L_i$ given by
\begin{equation}\label{eq:Li}
	L_i =  \!\int_{\Omega_{i}}\!\!\mathcal{L}^{\rm MHD}\d V + \tau_i(S_i - S_{i0}) +\mu_i\left(K_i - K_{i0}\right) \, .
\end{equation}
Here $\Omega_{i}$ denotes a plasma relaxation region and $\mathcal{L}^{\rm MHD}$ is the standard MHD Lagrangian density \cite{Newcomb_62,Dewar_70}, $\rho v^2/2 - p/(\gamma-1) - B^2/(2\muSI)$, with $\rho$ denoting mass density, $p$ the plasma pressure, and $\gamma$ the ratio of specific heats. The departure from ideal MHD is in constraining \emph{total} entropy $S_i$ and magnetic helicity $2\muSI K_i$ in each macroscopic  subregion $\Omega_i$, rather than in each microscopic fluid element $\d V$ (though mass is still microscopically conserved in the currrent 
formulation), the nonholonomic conservation of $S_i$ and $K_i$ being enforced through the Lagrange multipliers $\tau_i$ and $\mu_i$, respectively.

The entropy and magnetic helicity invariants are given explicitly by
\begin{equation}\label{eq:Sdef}
	S_i \equiv \int_{\Omega_i}\frac{\rho}{\gamma - 1} \ln\left(\kappa\frac{p}{\rho^\gamma}\right) \d V \;,
\end{equation}
and
\begin{equation}\label{eq:Helicity}
	K_i \equiv \int_{\Omega_i} \frac{\vrm{A}\dotv\vrm{B}}{2\muSI} \, \d V \;,
\end{equation}
where $\vrm{A}$ is a vector potential for $\vrm{B}$ with gauge arbitrary save for the constraint that loop integrals $\oint\!\vrm{A}\dotv\d\vrm{l}$ on the boundary and interface be conserved.
The constant $\kappa$ in \eqref{eq:Sdef}, required to make the argument of $\ln$ dimensionless, is arbitrary for our purposes but is identified physically in Appendix~A of \cite{Dewar_Yoshida_Bhattacharjee_Hudson_15}.
The constant reference values $S_{i0}$ and $K_{i0}$ are the respective initial values at $t = t_0$ evaluated over $\Omega_{i0}$, making $L_i = L^{\rm MHD}_i$ when the $\tau_i(t)$ and $\mu_i(t)$ are determined so as to satisfy conservation of $S_i$ and $K_i$.

Microscopic mass conservation is ensured by the continuity equation
\begin{equation}\label{eq:rho}
	\frac{\partial\rho}{\partial t} = -\divv(\rho\vrm{v}) \;,
\end{equation}
and the other equations of the model are found in \cite{Dewar_Yoshida_Bhattacharjee_Hudson_15} as Euler--Lagrange equations making the action $\int L\,\d t$ stationary under arbitrary Eulerian variations of the free fields $p$ and $\vrm{A}$, and variations arising from infinitesimal displacements of the Lagrangian positions of the fluid elements (including geometrical variation of the boundaries $\partial\Omega_i$).

Varying $p$ we find that the pressure within $\Omega_i$ obeys an \emph{isothermal} equation of state
\begin{equation}\label{eq:isoeqstate}
	p = \tau_i \rho \;,
\end{equation}
with the Lagrange multiplier $\tau_i$ thus identified as the spatially constant \emph{specific temperature} $T_i/M$ in $\Omega_i$, where $M$ is the effective ion mass $m_{\rm i}/Z_{\rm eff}$, $Z_{\rm eff}$ being the mean ionization state. An alternative physical interpretation of $\tau_i$ is as $C^2_{{\rm s}i}$, where $C_{{\rm s}i}$ is the isothermal \emph{sound speed} in $\Omega_i$.

Variation of $\vrm{A}$ gives the \emph{Beltrami equation},
\begin{equation}\label{eq:Beltrami}
	\curl\vrm{B} = \mu_i \vrm{B} \;.
\end{equation}

Variation of fluid element positions \emph{within} $\Omega_i$ gives the equation of motion
\begin{equation}\label{eq:momentumeq}
	\rho\left(\frac{\partial\vrm{v}}{\partial t} + \vrm{v}\dotv\grad\vrm{v}\right) = -\grad p \;,
\end{equation}
no contribution from the Lorentz force appearing, consistently with the fact that \eqref{eq:Beltrami} describes a ``force-free''  ($\vrm{j}\cross\vrm{B} = 0$) field in region $\Omega_i$.

Variation of fluid positions \emph{at} the interface $\partial\Omega_{i,j} \equiv \partial\Omega_{i}\cap\partial\Omega_{j}$ gives the force balance condition across the current sheet on this boundary
\begin{equation}\label{eq:normsurfvar}
	\jump{p + \frac{B^2}{2\muSI}}_{i,j} = 0 \;,
\end{equation}
the brackets $\jump{\cdot}_{i,j}$ denoting the jump in a quantity as the observation point crosses the interface from the $\Omega_{i}$ side to the $\Omega_{j}$ side.

To complete the specification of the MRxMHD equations we give the boundary conditions on $\vrm{B}$, namely tangentiality on region boundaries
\begin{equation}\label{eq:Bn}
	\vrm{n}_i\dotv\vrm{B} = 0 \:\: \text{on} \:\:\partial\Omega_i \;,
\end{equation}
and on $\vrm{v}$, normal continuity across moving (advected) interfaces
\begin{equation}\label{eq:vn}
	\vrm{n}_i\dotv\jump{\vrm{v}}_{i,j} = 0 \:\: \text{on} \:\:\partial\Omega_{i,j} \;,
\end{equation}
where $\vrm{n}_i$ is the outward unit normal at each point on $\partial\Omega_i$. Equation~\eqref{eq:vn} includes the case of a perfectly conducting confining wall, $j=\mathrm{w}$, moving with velocity $\vrm{v}_{\rm w}$, but the case of $\Omega_j$ being a vacuum region is obviously exceptional, as velocity is not defined in a vacuum. In this case there is no constraint on $\vrm{n}_i\dotv\vrm{v}_{i}$.

Finally, we define the \emph{normal flow velocity} $\vrm{v}_{\rm n}(\vrm{x},t) \equiv v_{\rm n}\vrm{n}$ where $v_{\rm n} \equiv \vrm{n}_i\dotv\vrm{v}$ is the normal component of the full fluid velocity $\vrm{v}$. Unlike $\vrm{v}$, $\vrm{v}_{\rm n}$ is [by \eqref{eq:vn}] the same on both sides of the interface, and thus provides a suitable generator for describing the geometric evolution of the interface $\partial\Omega_{i,j}$:  we define the \emph{normal flow map} $\vrm{r}^t_{\rm n}(\vrm{x}|t_0):\partial\Omega_{i,j}^t \to \partial\Omega_{i,j}^{t_0}$ by following the loci of points $\vrm{r}^t_{\rm n}(\vrm{x}|t_0)$ obeying $\d_{\rm n}\vrm{r}^t_{\rm n}/\d t \equiv (\partial_t + v_{\rm n}\vrm{n}\dotv\grad)\vrm{r}_{\rm n}^t$, $\vrm{r}^{t_0}_{\rm n}(\vrm{x}|t_0) \equiv \vrm{x}$ [cf. the Lagrangian flow map defined through eq.~(3.1) of \cite{Dewar_Yoshida_Bhattacharjee_Hudson_15}]\footnote{The interface evolution is defined in terms of the \emph{normal} flow velocity to remove secular growth in $|t - t_0|$ from the map when there is fluid flow tangential to the interface}. This map is well defined provided the interface is sufficiently smooth that $\vrm{n}$ is uniquely defined at each point on the interface (i.e. it must have no cusp-like behavior).

Note that, as in ideal MHD, there is no dissipation in MRxMHD. Unlike ideal MHD, the fluid and magnetic field are decoupled except at the interfaces, but as the number of interfaces may be arbitrarily large an arbitrarily high degree of coupling transverse to the interfaces can in principle be obtained. Extra physics, such as dissipation and current drive (helicity injection) can be added after the equations have been derived, but are not inherent in the theory proper. 

\section{Linearized equations}\label{sec:linGen}

We now suppose the solutions to the equations above can be written as the sum of a flowless, time-independent equilibrium part and arbitrarily small perturbations, $\vrm{B} = \vrm{B}_0 + \vrm{B}_1 = \curl(\vrm{A}_0 + \vrm{A}_1)$, $\vrm{v} = \vrm{v}_1$, $ \rho = \rho_0 + \rho_1$, and $p = p_0 + p_1$ where
\begin{equation}\label{eq:p1}
	p_1 = \tau_0\rho_1 + \tau_1\rho_0 \;.
\end{equation}
The linearized versions of \eqref{eq:rho} and \eqref{eq:momentumeq} are
\begin{equation}\label{eq:rho1}
	\frac{\partial\rho_1}{\partial t} = -\rho_0\divv\vrm{v} \;, 
\end{equation}
\begin{equation}\label{eq:momentumeq1}
	\rho_0\frac{\partial\vrm{v}}{\partial t} = -\tau_0\grad \rho_1 \;, 
\end{equation}
$\grad\rho_0$ being zero in a flowless relaxed equilibrium.

The linearized version of \eqref{eq:Beltrami} is
\begin{equation}\label{eq:Beltrami1}
	\curl\vrm{B}_1 - \mu_{i0}\vrm{B}_1 = \mu_{i1}\vrm{B}_0 \;,
\end{equation}
Unlike $\vrm{v}$, which obeys the evolution equation \eqref{eq:momentumeq1}, $\vrm{B}_1$ has no evolution equation. Instead it must be found at each instant in time by solving the inhomogeneous elliptic partial differential equation \eqref{eq:Beltrami1} (PDE) in each unperturbed region $\Omega_{i0}$ under appropriate boundary conditions. These are the linearizations of the tangentiality constraint \eqref{eq:Bn} and the loop integral constraints $\oint\!\vrm{A}\dotv\d\vrm{l} = \const$ on the disjoint components of $\partial\Omega_{i0}$, which are succinctly captured by the vector potential constraint (see e.g. Appendix B of \cite{Dewar_Yoshida_Bhattacharjee_Hudson_15})
\begin{equation}\label{eq:abc}
	\vrm{n}_0\cross(\vrm{A}_1 - \xi_{\rm n}\vrm{n}_0\cross\vrm{B}_0 -  \grad\chi_1) = 0 \:\:\text{on}\:\: \partial\Omega_{i0} \;, 
\end{equation}
where $\vrm{n}_0(\vrm{x})$ is the unit normal on the unperturbed interface and 
$\chi_1$ is an arbitrary single-valued gauge potential.

We decompose the solution of \eqref{eq:Beltrami1} as
\begin{equation}\label{eq:bdecomp}
	\vrm{B}_1 = \curl\vrm{a}_{\xi} + \frac{\mu_{i1}}{\mu_{i0}}\vrm{G}_0 \;,
\end{equation}
where $\vrm{G}_0(\vrm{x},t) \equiv \curl\vrm{A}_{{\rm G}0}$ is the solution in $\Omega_{i0}$ of the inhomogeous equation
\begin{equation}\label{eq:BeltramiG}
	\curl\vrm{G}_0 - \mu_{i0}\vrm{G}_0 = \mu_{i0}\vrm{B}_0 \;,
\end{equation}
with homogenous boundary condition $\vrm{A}_{{\rm G}0} = 0$ on $\partial\Omega_{i0}$; and $\curl\vrm{a}_{\xi}$ is the solution of the homogeneous equation
\begin{equation}\label{eq:Beltramiaxi}
	\curl(\curl\vrm{a}_{\xi}) - \mu_{i0}\curl\vrm{a}_{\xi} = 0
\end{equation}
driven by the inhomogeneous boundary condition \eqref{eq:abc} with $\vrm{A}_1 \mapsto \vrm{a}_{\xi}$.
While the general construction of this solution is rather complicated, our slab test case is sufficiently simple that we will be able to construct the solution using elementary methods.

To apply the boundary condition \eqref{eq:normsurfvar} on the perturbed interface $\partial\Omega_{i,j}$, we define the (linear) normal displacements $\xi_{\rm n}$ of points on $\partial\Omega_{i,j}$ away from their positions on the equilibrium interface $\partial\Omega^{(0)}_{i,j}$ through the normal flow map defined at the end of the previous section,
\begin{equation}
	\xi_{\rm n}(\vrm{x},t) \equiv \vrm{n}_0\dotv\vrm{r}^t_{\rm n}(\vrm{x}|-\infty) \;.
\end{equation}
In this equation it is assumed that a stable perturbation was switched on adiabatically from $t = -\infty$, or an unstable perturbation has grown from a negligible amplitude in the distant past, so $\vrm{r}^t_{\rm n}$ is essentially independent of $t_0$. Then the Eulerian linearization of \eqref{eq:normsurfvar} is
\begin{equation}\label{eq:normsurfvar1}
	\jump{p_1 + \frac{\vrm{B}_0\dotv\vrm{B}_1}{\muSI}}^{(0)}_{i,j} 
	+ \xi_{\rm n}\jump{\vrm{n}_0\dotv\grad\left(p_0 + \frac{B_0^2}{2\muSI}
	\right)}^{(0)}_{i,j} = 0 \;,
\end{equation}
respectively, and $\jump{\cdot}^{(0)}$ denotes a jump evaluated on $\partial\Omega^{(0)}_{i,j}$ (the branches of functions defined in the two adjacent regions $\Omega_i$ and $\Omega_j$ being assumed differentiable, so they may be extended, at least to linear order, into a neighbourhood on either side of $\partial\Omega_{i,j}$).

The normal component of the Eulerian perturbation in $\vrm{B}$ is given, \cite[eq.~(107)]{Spies_Lortz_Kaiser_01} and \cite{SuppCitePPCF}, in terms of $\xi_{\rm n}$ as
\begin{equation}\label{eq:B1} 
	\vrm{n}_0\dotv\vrm{B}_1 = \vrm{B}_0\dotv\grad\xi_{\rm n} + \xi_{\rm n}\vrm{n}_0\dotv
	\curl(\vrm{n}_0\cross\vrm{B}_0) \;.
\end{equation}

The linear perturbation to the entropy, \eqref{eq:Sdef}, is \cite{SuppCitePPCF}
\begin{equation}\label{eq:S1}
		S_{i1} =
		\rho_{i0}\int_{\partial\Omega_{i0}}\xi_{\rm n}\d S
		+ \frac{\tau_{i1}}{\tau_{i0}}\frac{\rho_{i0} V_{i0}}{\gamma - 1} \;,
\end{equation}
where we have used the mass conservation identity \cite{SuppCitePPCF} 
\begin{equation}\label{M1}
	\int_{\partial\Omega_{i0}}\rho_0\xi_{\rm n}\d S + \int_{\Omega_{i0}}\rho_1\d V = 0 \;.
\end{equation}
Setting $S_{i1} = 0$ to satisfy the entropy constraint gives
\begin{equation}\label{eq:tau1}
	\frac{\tau_{i1}}{\tau_{i0}} = -(\gamma - 1)\frac{V_{i1}}{V_{i0}} \;,
\end{equation}
where $V_{i0} \equiv \int_{\Omega_{i0}}\d V$ and $V_{i1} \equiv \int_{\partial\Omega_{i0}}\xi_{\rm n}\d S$.
In the case of spatially constant $\rho_1$ this is seen to be consistent with the adiabatic law $pV^\gamma = \const$, but is more general as \eqref{eq:tau1} does not require $\rho_1$ to be constant.

The linear perturbation to the magnetic helicity functional, \eqref{eq:Helicity} times $\muSI$, is
\begin{equation}\label{eq:K1}
	\muSI K_{i1}
	= \int_{\Omega_{i0}} \left(\vrm{B}_0\dotv\vrm{a}_{\xi} + \frac{\mu_{i1}}{\mu_{i0}}\vrm{A}_0\dotv\vrm{G}_0\right)\, \d V \;,
\end{equation}
where we have eliminated $\xi_{\rm n}$ using the boundary identity $\vrm{B}_0 \, \xi_{\rm n} =  \vrm{a}_\xi\cross\vrm{n}_0 + \vrm{n}_0\cross\grad\chi_1$, following from \eqref{eq:abc}\footnote{The gauge term is included for completeness, but does not contribute as a surface integration by parts converts it to $\chi_1\vrm{n}_0\dotv\vrm{B}_0$, which vanishes because of the tangentiality condition.}, and used the decomposition \eqref{eq:bdecomp} \cite{SuppCitePPCF}. 

Setting $K_{i1} = 0$ to satisfy the helicity constraint gives
\begin{equation}\label{eq:mu1}
	\frac{\mu_{i1}}{\mu_{i0}} = -\frac{\int_{\Omega_{i0}} \vrm{B}_0\dotv\vrm{a}_\xi \, \d V}
								{\int_{\Omega_{i0}} \vrm{A}_0\dotv\vrm{G}_0\, \d V} 
								\;.
\end{equation}

Except in the special case of purely radial waves discussed in Sec.~\ref{sec:mnzero} we shall find that $\tau_1$ and $\mu_1$ vanish.

\section{Topologically toroidal slab}\label{sec:slabgeom}

\subsection{Core-plus-slab model}\label{sec:coreslab}

As sketched in Fig.~\ref{fig:CylSlab} we view the plasma slab as a model, albeit imperfect, of the outer region of a toroidally confined plasma in which the two slab regions $\Omega^\sglt$ are regarded as topologically toroidal volumes (specifically, \emph{annular} toroids) surrounding a cylindrical core of relaxed plasma containing the ``magnetic axis'' at $r = 0$, a closed field line on which the poloidal angle $\theta$ is singular. 

In the following we constrain the core-slab boundary $r = a$ to be rigid, so that all \emph{dynamics} occurs within the slab regions. However, we shall take into account the core plasma in setting up the slab \emph{equilibrium} by assuming continuity of the rotational transform $\iotabar_{\rm c}(r)$, or equivalently its inverse $q_{\rm c}(r)$ [see \eqref{eq:qdef}], across the core-slab interface. Assuming the plasma within the cylindrical core obeys the Beltrami equation \eqref{eq:Beltrami}, with constant $\mu_{\rm c}$, the solution of \eqref{eq:Beltrami} is well known to be obtainable in Bessel functions (see e.g. \cite[eq.~(27)]{Hole_Hudson_Dewar_07}). The corresponding $q$-factors at the magnetic axis $r = 0$ and core edge $r = a$, denoted $q_{\rm c}(0)$ and $q_{\rm c}(a)$ respectively, are then found to be
\begin{equation}\label{eq:cylqs}
	q_{\rm c}(0) = \frac{2}{R \mu_{\rm c}}\;, \quad q_{\rm c}(a) = \frac{a J_0(a\mu_{\rm c})}{R J_1(a\mu_{\rm c})} \;,
\end{equation}
where the major radius $R$ of the torus which the cylinder approximates imparts the periodicity length $2\pi R$ in the $z$-direction. 

The magnetic axis cannot be included in our slab model because a slab region has no coordinate singularity. Thus we take the origin of the slab radial coordinate $x$ to be the core-slab boundary, so $r = a + x$. 

\subsection{Slab equilibrium}\label{sec:slabeqm}

The slab coordinates $y$ and $z$ are analogues of poloidal and toroidal coordinates: the poloidal and toroidal angles are given by $\theta \equiv y/a$ and $\zeta \equiv z/R$, respectively, where $R$ is the nominal major radius and $a$ is the minor radius of the notional core. Thus the periodicity lengths in the $y$ and $z$ directions are $L_{\rm pol} = 2\pi a$ and $L_{\rm tor} = 2\pi R$, respectively. For example, the unperturbed volumes of the plasma domains $\Omega^\sglt_0$ are 
\begin{equation}\label{eq:V0}
\begin{split}
	V^\sglt_0 &= \int_{x^\sglt_{-}}^{x^\sglt_{+}}\!\!\d x\int_0^{2\pi a}\!\!\!\!\d y\int_0^{2\pi R}\!\!\!\!\d z  \\
		&= (2\pi)^2 a a_{\rm i}^\sglt R \;,
\end{split}
\end{equation}
where, within the regions $\Omega_0^\sglt$, we have used \emph{subscripts} ${}_+$ and ${}_-$ to denote their outer and inner boundaries, respectively, and the notation $a_{\rm i}^\sglt$ for the widths of these regions:
\begin{equation}\label{eq:adef}
\begin{split}
	x^\slt_{-} &= 0\;, \:\: x^\slt_{+} \equiv x^\sgt_{-} = x_{\rm i}\;, \:\: \text{and}\:\: x^\sgt_{+} = x_{\rm w} \;, \\
	a_{\rm i}^\slt &\equiv x^\slt_{+} - x^\slt_{-} = x_{\rm i}\;, \:\:  a_{\rm i}^\sgt \equiv x^\sgt_{+} - x^\sgt_{-} = x_{\rm w} - x_{\rm i} \;.
\end{split}
\end{equation} 

We consider a two-layer slab equilibrium in which the inner region $\Omega^\slt_0: x^\slt_{-} = 0 \leq x < x^\slt_{+}\equiv x_{\rm i}$ is a region of relaxed plasma, with magnetic field $\vrm{B}^\slt(x)$. The outer region $\Omega^\sgt_0: x^\sgt_{-}\equiv x_{\rm i} < x \leq x^\sgt_{+} \equiv x_{\rm w}$ may either be a vacuum region in which the unperturbed magnetic field is a spatially constant vector $\vrm{B}_{\rm v}$, or a region of relaxed plasma, with magnetic field $\vrm{B}^\sgt(x)$. We assume no equilibrium flow: $\vrm{v}^\slt_0 = \vrm{v}^\sgt_0 = 0$.

In the following we use the notation ${\scriptsize\gtrless}$ to denote ``$>$ and/or $<$'' in an analogous way to how $\pm$ denotes ``$+$ and/or $-$'' (similarly ${\scriptsize\lessgtr}$ is the analogue of $\mp$). When we wish to associate a $+$ sign with $>$ and a $-$ sign with $<$ we extend the domain of the standard sign function $\sgn(\cdot)$ by defining $\signgt \equiv +1$, $\signlt \equiv -1$, $\signglt \equiv \pm 1$ and $\signlgt \equiv -\signglt = \mp 1$.

In relaxed plasma domains we take the unperturbed equilibrium fields to be, consistently with \eqref{eq:Beltrami}, force-free solutions of the form
\begin{equation}\label{eq:Beltrami0}
	\vrm{B}^\sglt_0(x) = B^\sglt_0\left[\esub{y}\sin(\Theta_0^\sglt + \mu^\sglt_0 x) + \esub{z}\cos(\Theta_0^\sglt + \mu^\sglt_0 x)\right] \;.
\end{equation}
The field magnitudes $B^\sglt_0$ are spatially constant, as are the angles $\Theta_0^\sglt$. Projected onto the $z,y$ plane the field lines form two families of planar magnetic surfaces parametrized by $x$, the field lines subtending angles $\Theta_0^\sglt + \mu^\sglt_0 x$ with the $z$-axis. Thus $\Theta_0^\slt$ is the angle subtended by field lines on the inner boundary, $x = x^\slt_{-} = 0$, of $\Omega^\slt$, whereas $\Theta_0^\sgt$ is the \emph{extrapolation} to $x = 0$ of angles subtended by field lines in $\Omega^\sgt$. As $x$ increases, the field lines rotate counterclockwise in the $z,y$ plane for $\mu^\sglt_0 > 0$, clockwise for $\mu^\sglt_0 < 0$, at constant shear rates determined by $|\mu^\sglt_0|$.

\begin{figure}[htbp]
   \centering
		\includegraphics[width = 0.9\textwidth]{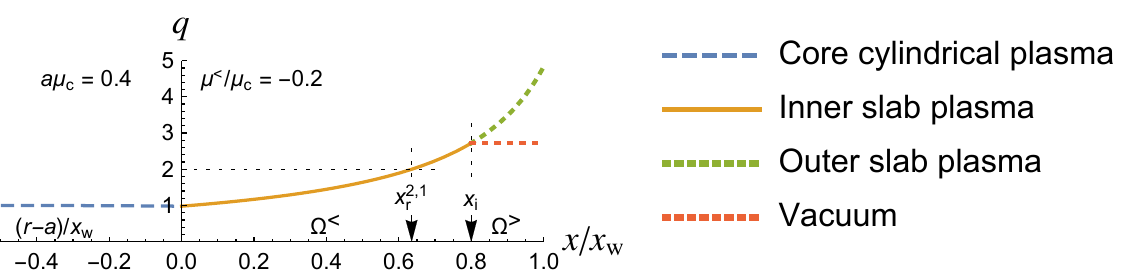} 
\caption{Typical tokamak-like $q$-profiles for a Taylor-relaxed cylindrical core ($x < 0$, dashed blue curve) + two-region MRxMHD slab model with interface at $x_{\rm i} = 0.8 x_{\rm w}$ separating an inner relaxed slab plasma region ($0 \leq x < x_{\rm i}$, solid orange curve), and either an outer plasma region ($x_{\rm i} < x < x_{\rm w}$, short-dashed green curve) or a vacuum region ($x_{\rm i} < x < x_{\rm w}$, short-dashed red line). See Appendix B of \cite{SuppCitePPCF} for details (colour coding online).} 
\label{fig:qprofTok}
\end{figure}

Continuing the analogy with tokamaks and other toroidal plasma confinement devices we define, as a measure of the winding number or helical pitch of the magnetic field lines, the ``safety factor'' $q(x)$ given by
\begin{equation}\label{eq:qdef}
	q(x) \equiv \frac{a B_{0z}(x)}{R B_{0y}(x)} = \frac{\epsilon_{\rm a} B_{0z}(x)}{B_{0y}(x)} \;,
\end{equation}
where  $\epsilon_{\rm a} \equiv a/R$ is the \emph{inverse aspect ratio}. [In stellarators its inverse, the \emph{rotational transform} $\iotabar(x) = 1/q(x)$ is often used instead.]

Using \eqref{eq:Beltrami0} in \eqref{eq:qdef} we find
\begin{equation}\label{eq:qglt}
	q^\sglt(x) = \epsilon_{\rm a}\cot(\Theta_0^\sglt + \mu^\sglt_0 x)
\end{equation}
In the case of a vacuum field in the outer slab region $\Omega^\sgt$, $\mu_0^\sgt = 0$. From \eqref{eq:qglt} this implies $q^\sgt(x)$ is a constant, $q_{\rm v} \equiv \epsilon_{\rm a}\cot(\Theta_0^\sgt)$. 

\begin{figure}[htbp]
   \centering
		\includegraphics[width = 0.95\textwidth]{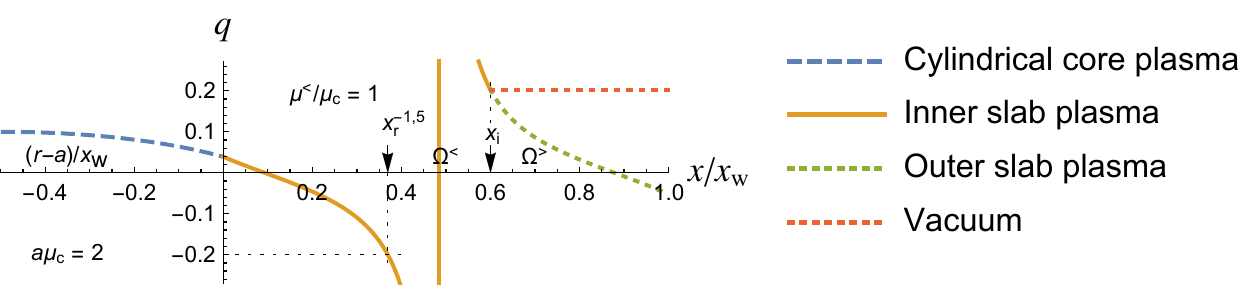} 
\caption{Typical pinch-like $q$-profiles for the a similar two-region slab model as in Fig.~\ref{fig:qprofTok}. Parameter values are indicated on the plot and in Appendix B of \cite{SuppCitePPCF}.} 
\label{fig:qprofRFP}
\end{figure}

While the radial profiles of density, pressure, and current are important for understanding modes and instabilities in a toroidal plasma experiment, arguably the most important radial profile is $q(r)$. 
Adopting the categorization of toroidal equilibria used by \cite{Bhattacharjee_Dewar_etal_83} into \emph{tokamak-like} ($q \gtrsim 1$ and increasing outward) and \emph{pinch-like} ($q \ll 1$ and initially decreasing outward) we give an example of each type of equilibrium 
in Figs.~\ref{fig:qprofTok} and \ref{fig:qprofRFP}. Details on the construction of these cases are given in Appendix B of \cite{SuppCitePPCF}. 

The magnitudes of the magnetic fields on either side of the equilibrium interface $x = x_{\rm i}$ separating the inner plasma from the outer plasma, or a zero-pressure vacuum, are related by equilibrium force balance. That is, from \eqref{eq:normsurfvar},
\begin{equation}\label{eq:normsurfvar0}
\begin{split}
		B^\slt_0(1 + \beta^\slt_0)^{1/2} &= (1 + \beta^\sgt_0)^{1/2}B^\sgt_0 \;,
\end{split}
\end{equation}
where the constants $\beta^\sglt_0 \equiv 2\muSI p^\sglt_0/B^{\sglt 2}_0$ are the ratios of the equilibrium plasma and magnetic pressures in the two regions. Later we shall also find it useful to write $\beta$ in the form
\begin{equation}\label{eq:betaalt}
	\beta = 2\left(C_{\rm s}/v_{\rm A}\right)^2 \;,
\end{equation}
where $C_{\rm s} \equiv p/\rho$ is the sound speed [cf. \eqref{eq:isoeqstate} ff.] and $v_{\rm A} \equiv B/\sqrt{\muSI\rho}$ is the \emph{Alfv\'en speed}.

Equation \eqref{eq:normsurfvar0} implies that, in the case of a finite pressure differential across the interface, there must necessarily be a jump in $|\vrm{B}|$, and hence a current sheet on the interface. Even if the pressure is continuous there may still be a current sheet if there is a \emph{tangential discontinuity} in $\vrm{B}$, i.e. if $q^\slt_{\rm i} \neq q^\sgt_{\rm i}$.

\section{Plane wave perturbations}\label{sec:slabeqwaves}

\subsection{Equilibrium variations}\label{sec:slabeqpert} 

To model purely radial modes we can use mass, entropy, magnetic flux, and magnetic helicity conservation to calculate perturbations in equilibrium plasma and magnetic field parameters under variations $\delta x_{\rm i}$.

Mass conservation implies $\delta\rho^\sglt_0/\rho^\sglt_0 = -\delta V^\sglt_0/V^\sglt_0 = -\delta a_{\rm i}^\sglt/a_{\rm i}^\sglt$, from \eqref{eq:V0}. From \eqref{eq:adef} we have
\begin{equation}\label{eq:avar}
		\delta a_{\rm i}^\sglt = -\signglt \delta x_{\rm i}  \;,
\end{equation}
where $\signglt \equiv \pm 1$ was defined in Sec.~\ref{sec:slabeqm}. 

Substituting \eqref{eq:V0} in \eqref{eq:tau1} we find the equilibrium temperature fluctuation from entropy conservation as 
$\delta\tau^\sglt_0/\tau^\sglt_0 = -(\gamma - 1)\delta a_{\rm i}^\sglt/a_{\rm i}^\sglt$, and combining with the results above we have the equilibrium pressure fluctuation as that expected from the ideal gas law,
\begin{equation}\label{eq:pvar}
	\frac{\delta p^\sglt_0}{p^\sglt_0} = -\gamma\frac{\delta a_{\rm i}^\sglt}{a^\sglt} \;.
\end{equation}

To calculate magnetic helicity we need suitable vector potentials corresponding to the magnetic fields in \eqref{eq:Beltrami0}, 
\begin{equation}\label{eq:A0}
\begin{split}
	\vrm{A}^\sglt_0(x) 
		&=  \frac{\vrm{B}^\sglt_0(x) - \vrm{B}^\sglt_0(x_{\rm i})}{\mu^\sglt_0} \;,\:\: x^\sglt_{-} < x \leq x^\sglt_{+} \;,
\end{split}
\end{equation}
where we have 
assumed gauges 
such that $\vrm{A}^\sglt_0(x_{\rm i}) = 0$, which, as $x^\sglt_\mp = x_{\rm i}$, ensures that the line integrals of $\vrm{A}_0$ on each side of the interface are equal. (Because $\vrm{B}$ is finite, an  interface of zero width can carry no magnetic flux.)

The special case where $\vrm{B}^{\sgt}_0$ is the shearless equilibrium vacuum field $\vrm{B}_{{\rm v}0} = B^\sgt_0 (\esub{y}\sin\Theta_{\rm v} +Ê\esub{z}\cos\Theta_{\rm v})$ can be obtained by taking the limit $\mu^\sgt_0 \to 0$ in \eqref{eq:Beltrami0} in such a way that $\Theta^\sgt \to \Theta_{\rm v}$. Taking this limit in \eqref{eq:A0} gives\cite{SuppCitePPCF} the vacuum vector potential $\vrm{A}_{{\rm v}0} = \lim_{\mu^\sgt_0 \to 0} \vrm{A}^\sgt_0$ as
\begin{equation}\label{eq:A0vac} 
\begin{split}
	\vrm{A}_{{\rm v}0}(x) 
		&
		=  B^\sgt_0 [\esub{y}\,(x - x_{\rm i})\cos\Theta_{\rm v} 
		   - \esub{z}\,(x - x_{\rm i})\sin\Theta_{\rm v} ] \;. 
\end{split}
\end{equation}

The poloidal and toroidal fluxes trapped between the perfectly conducting boundaries of the two annular toroids $\Omega^\sglt$, and thus conserved, 
are differences between inner and outer line integrals $\oint\!\vrm{A}\dotv\d\vrm{l}$, specifically $\int_0^{2\pi a}[A^\sglt_{0y}(x^\sglt_{+}) - A^\sglt_{0y}(x^\sglt_{-})]\d y$ and $\int_0^{2\pi R}[A^\sglt_{0z}(x^\sglt_{+}) - A^\sglt_{0z}(x^\sglt_{-})]\d z$. As $\vrm{A}_0(x_{\rm i}) \equiv 0$, conservation of poloidal and toroidal fluxes thus require the boundary condition that $\vrm{A}_0(x^\sglt_ {\rm bdy})$ 
be a constant vector under variation of $x_{\rm i}$, where we have denoted the fixed boundary in each region as $x^\sglt_ {\rm bdy}$, defined by
\begin{equation}\label{eq:xrefdef}
	x^\slt_{\rm bdy} \equiv x^\slt_{-} = 0 \;, \quad x^\sgt_{\rm bdy} \equiv x^\sglt_{+} = x_{\rm w} \;,
\end{equation} 

From \eqref{eq:Beltrami0} and \eqref{eq:A0} the toroidal and poloidal flux constraints $\delta\vrm{A}_0(x^\sglt_ {\rm bdy}) = 0$ under variation of $x_{\rm i}$ (and consequent variations in $B^\sglt_0$, $\mu_0^\sglt$ and $\Theta_0^\sglt$) can be combined using complex exponential notation \cite{SuppCitePPCF}: 
\begin{subequations}
\begin{align}
	\delta\left\{\frac{B^\sglt_0}{\mu^\sglt_0} 
	\exp \ijot\!\left(\Theta_0^\sglt + \frac{\mu^\sglt_0(x^\sglt_{\rm bdy} + x_{\rmi})}{2}\right)
	 		\sin\frac{\mu^\sglt_0 a_{\rm i}^\sglt}{2} \right\} 
	&= 0 \;, \label{eq:gltflxbc} \\
\text{Which can be broken into three independent variational constraints:} & \nonumber \\
	 \frac{B^\sglt_0}{\mu^\sglt_0} \cos\!\left(\frac{\mu^\sglt_0 a_{\rm i}^\sglt}{2}\right) 
		\;\delta\!\left(\mu^\sglt_0 a_{\rm i}^\sglt\right)
	& = 0 \;,\label{eq:gltflxbc1} \\
	 	\sin \!\left(\frac{\mu^\sglt_0 a_{\rm i}^\sglt}{2}\right) 
		\;\delta\!\left(\frac{B^\sglt_0}{\mu^\sglt_0}\right)
	& = 0 \;,\label{eq:gltflxbc2} \\
		\left(\frac{B^\sglt_0}{\mu^\sglt_0}\right) \sin\! \left(\frac{\mu^\sglt_0 a_{\rm i}^\sglt}{2}\right)
		\;\delta\!\left[\Theta_0^\sglt + \frac{\mu^\sglt_0(x^\sglt_{\rm bdy} + x_{\rmi})}{2}\right]
	& = 0 \;. \label{eq:gltflxbc3}
\end{align}
\end{subequations}

Assuming $\mu^\sglt_0 \neq 0$, \eqref{eq:gltflxbc1} implies $\delta\!\left(\mu^\sglt_0 a_{\rm i}^\sglt\right) = 0$ [except possibly at the zeros of $\cos\left(\mu^\sglt_0 a_{\rm i}^\sglt/2\right)$], and  \eqref{eq:gltflxbc2} and \eqref{eq:gltflxbc3} imply $\delta\!\left(B^\sglt_0/\mu^\sglt_0\right) = 0$ and  $\delta\!\left[\Theta_0^\sglt + \mu^\sglt_0(x^\sglt_{\rm bdy} + x_{\rmi})/2\right] = 0$, respectively [except possibly at the zeros of $\sin(\mu^\sglt_0 a_{\rm i}^\sglt/2)$]. These constraints are sufficient to completely determine how $\mu^\sglt_0$, $B^\sglt_0$ and $\Theta_0^\sglt$ vary with $x_{\rm i}$.

In the vacuum field case $\mu_0^\sgt = 0$, we see from \eqref{eq:A0vac}, or taking the $\mu_0^\sgt \to 0$ limit of \eqref{eq:gltflxbc2} and \eqref{eq:gltflxbc3}, that there are now only two independent variational constraints in
$\Omega^\sgt$ \cite{SuppCitePPCF}, 
\begin{equation}\label{eq:vacflux} 
\begin{split}
	\delta\left[B^\sgt_0 (x_{\rm w} - x_{\rm i})\right] &= 0 \;, \\
	\delta\Theta_{\rm v} &= 0 \;, \\
\end{split}
\end{equation}
so $\Theta_{\rm v}$ is constant and $B^\sgt_0$ varies as expected from elementary flux conservation considerations.

We must now consider the constraints provided by the helicity integrals \eqref{eq:Helicity}. Equation~\eqref{eq:A0} gives (for $\mu_0 \neq 0$, which is all that is required as helicity conservation does not apply for a vacuum field),
\begin{equation}\label{eq:A0dotB0}
	\vrm{A}^\sglt_0\dotv\vrm{B}^\sglt_0(x) 
	= \frac{B^{\sglt 2}_0 - \vrm{B}^\sglt_0(x_{\rm i})\dotv\vrm{B}^\sglt_0(x)}{\mu^\sglt_0} \;, 	 \quad x^\sglt_{-} < x \leq x^\sglt_{+} \;.
\end{equation}
Multiplying by $2\muSI$ and using \eqref{eq:Beltrami0} in \eqref{eq:A0dotB0} the helicity constraints become \cite{SuppCitePPCF} 
\begin{equation}\label{eq:intA0B0}
\begin{split}
	&\int_{\Omega^\sglt_0} \vrm{A}^\sglt_0\dotv\vrm{B}^\sglt_0 \,\d V  \\
	\quad &= \frac{(2\pi)^2 a R B^{\sglt 2}_0}{\mu^{\sglt 2}_0}
	(\mu^\sglt_0 a_{\rm i}^{\sglt} - \sin\mu^\sglt_0 a_{\rm i}^{\sglt})  \;,
\end{split}
\end{equation}
except in the case of a vacuum field in the outer region, when only the result in $\Omega^\slt$ is relevant.
Conservation of magnetic helicity thus implies
\begin{equation}\label{eq:hel0constraint}
	\delta\left[\frac{B^{\sglt 2}_0}{\mu^{\sglt 2}_0}\left(\mu^\sglt_0 a_{\rm i}^\sglt - \sin \mu^\sglt_0 a_{\rm i}^\sglt\right)\right] = 0 \;,
\end{equation}
consistently with \eqref{eq:gltflxbc1} and \eqref{eq:gltflxbc2} and showing that the zeros of $\cos\left(\mu^\sglt_0 a_{\rm i}^\sglt/2\right)$ and $\sin\left(\mu^\sglt_0 a_{\rm i}^\sglt/2\right)$ are \emph{not} exceptional points.

Thus, in summary, we have shown \cite{SuppCitePPCF}
\begin{subequations} 
\begin{align}
	\frac{\delta \mu^\sglt_0}{\mu^\sglt_0} &= -\frac{\delta a_{\rm i}^{\sglt}}{a_{\rm i}^{\sglt}} \;, \label{eq:muvar} \\
	\frac{\delta\Theta_0^\sglt}{\mu_0^\sglt x^\sglt_{\rm bdy}} &= \;\frac{\delta a_{\rm i}^{\sglt}}{a_{\rm i}^{\sglt}} \label{eq:Thetavar} \;, \\
	\frac{\delta B^\sglt_0}{B^\sglt_0} &= -\frac{\delta a_{\rm i}^{\sglt}}{a_{\rm i}^{\sglt}} \label{eq:B0var} \;.
\end{align}
\end{subequations}
In the case of a vacuum field in the outer region, from \eqref{eq:vacflux} we see that \eqref{eq:B0var} remains correct in both inner and outer regions, \eqref{eq:muvar}$^\sgt$ is inapplicable and \eqref{eq:Thetavar}$^\sgt$ is replaced by $\delta\Theta_{\rm v} = 0$.

For linearised radial perturbations, even if time-dependent (waves), \eqref{eq:muvar} allows us to bypass the complicated construction of $\mu_1$ in \eqref{eq:mu1} as relaxation is assumed to effectively instantaneous in dynamical MRxMHD. Similarly, the expression for $\delta\tau_0^\sglt$ leading to \eqref{eq:pvar} may be used to get $\tau_1$ for radially propagating waves as temperature equilibration is assumed effectively instantaneous also. These results are used in Sec.~\ref{sec:mnzero}.

\subsection{Surface, plane and standing waves}\label{sec:slabpert}

Although the lower boundary $x = 0$ is notionally the inner face of an annular toroidal region within the plasma we assume it is not affected by wave perturbations and may be treated as a rigid boundary. Eigenmodes in this geometry are thus somewhat analogous to water waves \cite[\S 4.5]{Hosking_Dewar_15} in that the basic wave is a  transverse wave perturbation on the two-dimensional (2-D) interface $x = x_{\rm i}$, 
\begin{equation}\label{eq:surfwave}
	\xi_{\rm n}(y,z,t) = \Re\,\widetilde{\xi}\exp(\ijot\vrm{k}_{yz}\dotv\vrm{x} - \ijot\omega t) \;,
\end{equation}
where the 2-D wave vector $\vrm{k}_{yz} \equiv k_y\esub{y} + k_z\esub{z}$ and a tilde, such as in $\widetilde{\xi}$, denotes a complex amplitude. 

To make a correspondence with standard notations for waves in a toroidally confined plasma we write
\begin{equation}\label{eq:kmn}
\begin{split}
	\vrm{k}_{yz}^{m,n} &= \frac{m\esub{y}}{a} - \frac{n\esub{z}}{R} \\ 
					&= \frac{m\esub{y} - n \epsilon_{\rm a}\esub{z}}{a} \;,
\end{split}
\end{equation}
where $m$ is the poloidal and $n$ the toroidal mode number, and $\epsilon_{\rm a}$ is the inverse aspect ratio [see \eqref{eq:qdef}].

Associated with this surface wave are 3-dimensional (3-D) plane waves in $\Omega^{\sglt}$ of the generic form
\begin{equation}\label{eq:planewave}
	w^{\alpha\pm}(x,y,z,t) = \Re\,\widetilde{w}^{\alpha\pm}\exp(\ijot\vrm{k}^{\alpha\pm}\!\dotv\vrm{x} - \ijot\omega t) \;,
\end{equation}
where the scalar or vector $w$ is a member of the set of fields $\{\rho^\sglt_1,p^\sglt_1,\vrm{A}^\sglt_1,\vrm{B}^\sglt_1,\vrm{v}^\sglt \}$. 

The superscripts ${}^{\alpha\pm}$ are branch labels for the wave components contributing to the total 3-D response to the 2-D surface wave. The label $\alpha$ distinguishes two types of wave, sonic and magnetic (see Secs~\ref{sec:planesound} and \ref{sec:planemag} respectively), whose dispersion relations are both of the general form $D_{\alpha}(\omega^2, k^2) = 0$.
By definition $\omega$ is common to all components of an eigenmode, as is the 2-D wave vector $\vrm{k}_{yz}$. However, the 3-D wave vector,
\begin{equation}\label{eq:pwdecomp}
	\vrm{k}^{\alpha\pm} = k^{\alpha\pm}_x\esub{x} + \vrm{k}_{yz} \;,
\end{equation}
is branch-dependent through $k^{\alpha\pm}_x \equiv \pm k^{\alpha}_x$, where $k^{\alpha}_x$ is one of the two solutions of the local dispersion relation $D_{\alpha}\left(\omega^2, k^{\alpha 2}_x + |\vrm{k}_{yz}|^2\right) = 0$. While $\vrm{k}_{yz}$ is always real, $k^{\alpha}_x$ may be either real, corresponding to radial propagation, or imaginary, corresponding to radial evanescence. The details of this will be developed in Sec.~\ref{sec:eigen}.

Whether propagating or evanescent, the  plane waves are reflected from the inner and outer boundaries, $x = 0$ and $x_{\rm w}$, coupling the ${}^\pm$ branches to form radially standing waves. At the interface, 
the total perturbation is thus generically of the form
\begin{equation}\label{eq:standingwave}
		w^{\alpha} =\Re \widetilde{w}^{\alpha}_{\rm st}(x)\exp(\ijot\vrm{k}_{yz}\dotv\vrm{x} - \ijot\omega t)  \;,
\end{equation}
where
\begin{equation}\label{eq:wsw}
	\widetilde{w}^{\alpha}_{\rm st}(x) = \sum_{\pm}\widetilde{w}^{\alpha\pm}\!
		\exp(\pm\ijot k^{\alpha\pm}_x x) \;.
\end{equation}

The coefficients $\widetilde{w}^{\alpha\pm\sglt}$ are to be determined from the boundary conditions at the core-slab interface and the wall, 
\begin{equation}\label{eq:bcs}
\begin{split}
	v^\slt_x &= 0 \:\:\text{and}\: B^\slt_{1x} = 0 \:\: \text{on} \:\: x=0 \;,\\
	v^\sgt_x &= 0 \:\:\text{and}\: B^\sgt_{1x} = 0 \:\: \text{on} \:\: x=x_{\rm w} \;,
\end{split} 
\end{equation}
by \eqref{eq:Bn} and \eqref{eq:vn}; also those at the internal interface position, 
\begin{equation}\label{eq:ibcs}
	v_x(x_{\rm i},y,z,t) = \partial_t \xi_{\rm n}  \:\:\text{and}\: B^\sglt_{1x}(x_{\rm i},y,z,t) = \vrm{B}^\sglt_0\dotv\grad \xi_{\rm n}\;. 
\end{equation} 

The latter boundary condition follows from \eqref{eq:B1}, noting that $\vrm{n}_0\dotv\curl(\vrm{n}_0\cross\vrm{B}_0) = 0$ in slab geometry \cite{SuppCitePPCF}. In wave representation, \eqref{eq:ibcs} becomes
\begin{subequations}
\begin{align}
	v_x(x_{\rm i},y,z) &= \Re -\ijot\omega\,\widetilde{\xi}\exp(\ijot\vrm{k}_{yz}\dotv\vrm{x} - \ijot\omega t) \label{eq:vxw} \;,\\
	B^\sglt_{1x}(x_{\rm i},y,z,t) &= \Re\,\ijot\vrm{k}_{yz}\dotv\vrm{B}^\sglt_0\, \widetilde{\xi}\exp(\ijot\vrm{k}_{yz}\dotv\vrm{x} - \ijot\omega t) \;. \label{eq:bsurf}
\end{align}
\end{subequations}

In the general case $\vrm{k}_{yz} \neq 0$ (i.e. when at least one of $m$ and $n$ is nonzero), the average of $\exp(\ijot\vrm{k}_{yz}\dotv\vrm{x})$ over $y$ and $z$ vanishes, in which case we see from \eqref{eq:tau1} that $\tau_1 = 0$. Similarly the volume integral of $\vrm{B}_0\dotv\vrm{a}_\xi$ is zero, so, from \eqref{eq:mu1}, $\mu_1 = 0$. Thus we can take $\tau$ and $\mu$ to be constants when $\vrm{k}_{yz} \neq 0$ and suppress the subscript 0. (Even for $\vrm{k}_{yz} = 0$ we can easily calculate the effects associated with vanishing  $\tau_1$ and $\mu_1$ separately from finite-$\tau_1$ and $\mu_1$ effects and then linearly superpose them, as will be done in Sec.~\ref{sec:mnzero}.)

\subsection{Plane waves: acoustic wave local dispersion relation}\label{sec:planesound}

We now examine density-velocity waves propagating in the plasma, again suppressing the superscripts ${}^\sglt$ in this subsection as the results apply for both inner and outer plasmas (though not a vacuum). Using \eqref{eq:rho1} we find
\begin{equation}\label{eq:rho1F}
	 \widetilde{\rho}^{\pm} = \frac{\rho_0\vrm{k}^{{\rm s}\pm}\!\dotv\widetilde{\vrm{v}}^{\pm}}{\omega} \;.
\end{equation}
From \eqref{eq:momentumeq1} we have
\begin{equation}\label{eq:momentumeq1F}
	\widetilde{\vrm{v}}^{\pm} = \frac{\tau_0\widetilde{\rho}^{\pm} \vrm{k}^{{\rm s}\pm}}{\rho_0\omega} \;, 
\end{equation}
showing these waves are longitudinal.

Eliminating $\widetilde{\vrm{v}}^{\pm}$ by substituting \eqref{eq:momentumeq1F} in \eqref{eq:rho1F} we find the local dispersion relation for \emph{sound waves},
\begin{equation}\label{eq:sounddisprel}
	 \omega^2 = \tau_0 \vrm{k}^{{\rm s}2} \equiv C_{\rm s}^2\left[|\vrm{k}_{yz}|^2 + (k_x^{{\rm s}\pm})^2\right] \;,
\end{equation}
where $C_{\rm s} \equiv \sqrt{\tau_0}$ is the ion sound speed. We thus define $k_x^{{\rm s}\pm} \equiv \pm k_x^{\omega}$, where 
\begin{equation}\label{eq:ksound}
	 k_x^{\omega} \equiv \left(\frac{\omega^2}{C_{\rm s}^2} - |\vrm{k}_{yz}|^2\right)^{1/2}\;.
\end{equation}
The following expression for $\widetilde{\rho}^{\pm}$ in terms of $\widetilde{v}^\pm_x$, obtained from the $x$-component of \eqref{eq:momentumeq1F}, will be found useful when deriving eigenvalue equations,
\begin{equation}\label{eq:rho1Fx}
	 \widetilde{\rho}^{\pm} = \pm\frac{\rho_0\omega}{\tau_0 k_x^\omega}\,\widetilde{v}^\pm_x \;.
\end{equation}

As in ideal MHD, $\omega^2$ is real, so $\omega^2 < 0$ for all unstable modes (growth rate $\Im\omega > 0$). From \eqref{eq:sounddisprel} this implies the \emph{instability criterion} that $|k_x^{\omega}|^2 < -|\vrm{k}_{yz}|^2$. That is, $k_x^{\omega}$ must be located higher up the imaginary axis than $\ijot |\vrm{k}_{yz}|$,
\begin{equation}\label{eq:instability}
	 k_x^{\omega} = \ijot |k_x^{\omega}|, \:\: |k_x^{\omega}| > |\vrm{k}_{yz}| \;.
\end{equation}

\subsection{Plane waves: magnetic}\label{sec:planemag}

As well as suppressing the superscripts ${}^\sglt$, in this subsection we reduce the proliferation of subscripts by suppressing the subscript $0$ on equilibrium quantities and expressing the perturbed magnetic field, $\vrm{B}_1$ and vector potential $\vrm{A}_1$ as $\vrm{b}$ and $\vrm{a}$, respectively. As the case $\vrm{k}_{yz} = 0$ has already been treated in Sec.~\ref{sec:slabeqpert} we assume $\vrm{k}_{yz} \neq 0$, so $\mu_1 = 0$ and $\mu$ is the same constant value $\mu_0$ as used to construct the equilibrium in Sec.~\ref{sec:slabeqm}.

Following the general plan of Sec.~\ref{sec:slabpert} we introduce the plane-wave ansatz $\vrm{b}^{\pm} = \widetilde{\vrm{b}}^{\pm}\!\exp(\ijot\vrm{k}^{\mu\pm}\dotv\vrm{x} - \ijot\omega t)$, where [cf. \eqref{eq:pwdecomp}]
\begin{equation}\label{eq:pbwdecomp}
	\vrm{k}^{\mu\pm} \equiv k^{\mu\pm}_x\esub{x} + \vrm{k}_{yz} \;.
\end{equation}
Substituting this ansatz into the linearized Beltrami equation \eqref{eq:Beltrami1}, $\curl\vrm{b} = \mu \vrm{b}$, gives the plane-wave Beltrami equation, $\ijot\vrm{k}^{\mu\pm}\cross\widetilde{\vrm{b}}^{\pm} =  \mu\widetilde{\vrm{b}}^{\pm}\!\!.$ We now show this is solved by the ansatz \cite[Appendix C]{SuppCitePPCF} 
\begin{equation}\label{eq:bpm} 
\begin{split}
	\widetilde{\vrm{b}}^{\pm} & = 
	\widetilde{b}_x^{\pm}\left(\esub{x}
	- \frac{k_x^{\mu\pm}\vrm{k}_{yz}}{|\vrm{k}_{yz}|^2}
	+ \ijot \mu\frac{\vrm{k}_{yz}\cross\esub{x}}{|\vrm{k}_{yz}|^2}\right) \;, \\
\end{split}
\end{equation}
where $k_x^{\mu\pm} \equiv \pm k_x^{\mu}$, with $k_x^{\mu}$ defined by
\begin{equation}\label{eq:kxmu}
	\begin{split}
	& k_x^\mu = \left(\mu^2 - |\vrm{k}_{yz}|^2\right)^{1/2} \quad \text{for}\:\: |\mu| \geq |\vrm{k}_{yz}|   \;,\\
	& k_x^\mu = \ijot\!\left(|\vrm{k}_{yz}|^2 - \mu^2\right)^{1/2} \quad \text{for}\:\: |\mu| < |\vrm{k}_{yz}| \;,
	\end{split}
\end{equation}
the latter case corresponding to radially evanescent or growing waves. It is readily verified that both forms of $k_x^{\mu\pm}$ above satisfy
\begin{equation}\label{eq:rawdisprel}
	 (\vrm{k}^{\mu\pm})^2 \equiv (k_x^{\mu\pm})^2 + |\vrm{k}_{yz}|^2  = \mu^2 \;.
\end{equation}
[This ``local dispersion relation'' can be found by dotting each side of the ``Beltrami wave equation'' $\ijot\vrm{k}^{\mu\pm}\cross\widetilde{\vrm{b}}^{\pm} =  \mu\widetilde{\vrm{b}}^{\pm}\!\!$ with its complex conjugate. Unlike more usual dispersion relations its $\omega$-dependence is trivial (i.e. that of a constant) because the Beltrami equation has no time derivatives --- the relaxed magnetic field adjusts instantaneously to the boundary conditions.] 

Dotting \eqref{eq:bpm} with $\vrm{k}^{\mu\pm}$ gives $\vrm{k}^{\mu\pm}\dotv\widetilde{\vrm{b}}^{\pm} = 0$, the plane-wave version of $\divv\vrm{b} = 0$, verifying the magnetic perturbations are \emph{transverse}.
Dotting \eqref{eq:bpm} with itself gives $\widetilde{\vrm{b}}^{\mu\pm}\dotv\widetilde{\vrm{b}}^{\mu\pm} = 0$, showing that these transverse perturbations are \emph{circularly polarized} \cite{SuppCitePPCF} [similarly, e.g., to the complex unit vector $\widetilde{\vrm{e}} \equiv (\esub{x} + \ijot\esub{y})/\sqrt{2}$, which has the properties $\widetilde{\vrm{e}}\dotv\widetilde{\vrm{e}} = 0$, $\widetilde{\vrm{e}}^*\dotv\widetilde{\vrm{e}} = 1$].  
Also, multiplying both sides of \eqref{eq:bpm} by $\ijot$ and crossing with $\vrm{k}^{\mu\pm}$ gives \cite{SuppCitePPCF} 
\begin{equation}\label{eq:curlb}
	\ijot\vrm{k}^{\mu\pm}\cross\widetilde{\vrm{b}}^{\pm}  
	= \mu\widetilde{b}_x^{\pm}
	\left[\esub{x}
		- \frac{k_x^{\mu\pm}\vrm{k}_{yz}}{|\vrm{k}_{yz}|^2} 
	+\, \ijot\frac{(k_x^{\mu 2} + |\vrm{k}_{yz}|^2)\vrm{k}_{yz}\cross\esub{x}}{\mu|\vrm{k}_{yz}|^2}
	\right] \;, 
\end{equation}
which, using \eqref{eq:bpm} and \eqref{eq:rawdisprel}, reduces to $\ijot\vrm{k}^{\mu\pm}\cross\widetilde{\vrm{b}}^{\pm} =  \mu\widetilde{\vrm{b}}^{\pm}$ as required.

\section{Standing waves}\label{sec:standing}
\subsection{Standing sound waves}\label{sec:sound}

We now superpose the two ($\pm$) plane waves to give a radial standing wave [cf. \eqref{eq:standingwave}], giving $\vrm{v}^\sglt = \Re[\widetilde{\vrm{v}}^\sglt_{\rm st}(x) \exp(\ijot\vrm{k}_{yz}\dotv\vrm{x} - \ijot\omega t)]$ and $\rho^\sglt_1 = \Re[\widetilde{\rho}^\sglt_{\rm st}(x) \exp(\ijot\vrm{k}_{yz}\dotv\vrm{x} - \ijot\omega t)]$, where
\begin{equation}\label{eq:vsw}
\begin{split}
	\widetilde{\vrm{v}}^\sglt_{\rm st}(x) \equiv \sum_{\pm}\widetilde{\vrm{v}}^{\sglt\pm}\!\exp(\pm\ijot k_x^{\sglt\omega}x) \;, \\
	\widetilde{\rho}^\sglt_{\rm st}(x) \equiv \sum_{\pm}\widetilde{\rho}^{\sglt\pm}\!\exp(\pm\ijot k_x^{\sglt\omega}x) \;.
\end{split}
\end{equation}

To satisfy the velocity boundary conditions in \eqref{eq:bcs} we thus require $\esub{x}\dotv\widetilde{\vrm{v}}^\slt_{\rm st}(x^\slt_{-}) = \esub{x}\dotv\widetilde{\vrm{v}}^\sgt_{\rm st}(x^\sgt_{+}) = 0$ on the innermost boundary and the wall [using the notation of \eqref{eq:adef}], which we now show are satisfied by choosing amplitudes $\widetilde{v}_x^{\slt\pm} = \pm\widetilde{v}^\slt/2$ and $\widetilde{v}_x^{\sgt\pm} = \pm\widetilde{v}^\sgt\exp(\mp\ijot k_x^{\sgt\omega}x_{\rm w})/2$. 
From \eqref{eq:pwdecomp} and \eqref{eq:momentumeq1F} the $\widetilde{\vrm{v}}^ {\sglt\pm}$ are vectors in the $\vrm{k}^{{\rm s}\sglt\pm} = \pm k_x^{\sglt\omega}\esub{x} + \vrm{k}_{yz}$ direction. Normalizing to give the stated $x$-components and inserting in \eqref{eq:vsw} we find 
\begin{equation}\label{eq:vstandingwave}
\begin{split}
	\widetilde{\vrm{v}}^\slt_{\rm st} &= 
	\ijot\,\widetilde{v}^\slt\,\esub{x}\sin(k_x^{\slt\omega} x) 
	+ \widetilde{v}^\slt\,\frac{\vrm{k}_{yz}}{k_x^{\slt\omega}} \cos(k_x^{\slt\omega} x) \;, \\
	\widetilde{\vrm{v}}^\sgt_{\rm st} &= 
	-\ijot\,\widetilde{v}^\sgt\,\esub{x}\sin k_x^{\sgt\omega} (x_{\rm w} - x) 
	 + \widetilde{v}^\sgt\,\frac{\vrm{k}_{yz}}{k_x^{\sgt\omega}} \cos k_x^{\sgt\omega} (x_{\rm w} - x) 
	  \;,
\end{split}
\end{equation}
which indeed satisfies $\esub{x}\dotv\widetilde{\vrm{v}}^\slt_{\rm st}(0) = \esub{x}\dotv\widetilde{\vrm{v}}^\sgt_{\rm st}(x_{\rm w}) = 0$.

Setting $x = x_{\rm i}$ in \eqref{eq:vstandingwave} and comparing with the boundary condition \eqref{eq:vxw} allows us to relate $\widetilde{v}$ and $\widetilde{\xi}$,
\begin{equation}\label{eq:xivnreln}
\begin{split}
	 \widetilde{v}^\slt & = -\frac{\omega}{\sin(k_x^{\slt\omega} x_{\rm i})}\widetilde{\xi} \;,\\
	 \widetilde{v}^\sgt & = +\frac{\omega}{\sin k_x^{\sgt\omega} (x_{\rm w} - x_{\rm i})}\widetilde{\xi} \;,
\end{split}
\end{equation}
which can be summarized as $\widetilde{v}^\sglt = \pm  \omega\widetilde{\xi}/\sin (k_x^{\sglt\omega} a_{\rm i}^\sglt)$.

Using $\widetilde{v}_x^{\slt\pm} = \pm\widetilde{v}^\slt/2$ and $\widetilde{v}_x^{\sgt\pm} = \pm\widetilde{v}^\sgt\exp(\mp\ijot k_x^{\sgt\omega}x_{\rm w})/2$ in \eqref{eq:rho1Fx} gives, after using \eqref{eq:xivnreln},
\begin{equation}\label{eq:rhostandingwave}
	\widetilde{\rho}^\sglt_{\rm st}(x) 
	= \signglt\frac{\rho^\sglt_0\omega^2\cos (k_x^{\sglt\omega} |x - x^\sglt_{\sgn\!(\sglt)}|)}{\tau^\sglt_0 k_x^{\sglt\omega}\sin(k_x^{\sglt\omega} a_{\rm i}^\sglt)}\, \widetilde{\xi} \;,
\end{equation}
where $\signglt \equiv \pm 1$ was defined in Sec.~\ref{sec:slabeqm} and the subscript $\pm$ and $a_{\rm i}^\sglt$ notations were defined in \eqref{eq:adef}.

\subsection{Standing magnetic fluctuations}\label{sec:bplanewave}

Analogously to Sec.~\ref{sec:sound} we form (driven) standing magnetic ``Beltrami waves'' $\vrm{b}^\sglt = \Re[\widetilde{\vrm{b}}^\sglt_{\rm st}(x) \exp(\ijot\vrm{k}_{yz}\dotv\vrm{x} - \ijot\omega t)]$, where
\begin{equation}\label{eq:bsw}
	\widetilde{\vrm{b}}^\sglt_{\rm st}(x) = \sum_{\pm}\widetilde{\vrm{b}}^{\sglt\pm}\!
		\exp(\pm\ijot k_x^{\sglt\mu} x) \;.
\end{equation}

The boundary conditions $\esub{x}\dotv\widetilde{\vrm{b}}^\slt_{\rm st}(x^\slt_{-}) = \esub{x}\dotv\widetilde{\vrm{b}}^\sgt_{\rm st}(x^\sgt_{+}) = 0$ are satisfied by choosing amplitudes $\widetilde{b}_x^{\slt\pm} = \pm\widetilde{b}^\slt/2$ and $\widetilde{b}_x^{\sgt\pm} = \pm\widetilde{b}^\sgt\exp(\mp\ijot k_x^{\sgt\mu}x_{\rm w})/2$. Using \eqref{eq:bpm} we find
\begin{equation}\label{eq:bstandingwave}
\begin{split}
	\widetilde{\vrm{b}}^\sglt_{\rm st} &= 
	\ijot\widetilde{b}^\sglt\,\esub{x}\sin \left(k_x^{\sglt\mu} (x - x^\sglt_{\sgn\sglt})\right) 
	- \widetilde{b}^\sglt\frac{k_x^{\sglt\mu}\vrm{k}_{yz}}{|\vrm{k}_{yz}|^2} \cos \left(k_x^{\sglt\mu} (x - x^\sglt_{\sgn\sglt})\right)  \\
	& 
	- \mu^\sglt\widetilde{b}^\sglt\frac{\vrm{k}_{yz}\cross\esub{x}}{|\vrm{k}_{yz}|^2} \sin \left( k_x^{\sglt\mu} (x - x^\sglt_{\sgn\sglt})\right)   
	  \;,
\end{split}
\end{equation}
which 
indeed satisfies the required boundary conditions.

The remaining unknown, $\widetilde{b}$, is determined from the boundary condition \eqref{eq:bsurf}, which in the notation of this subsection is
\begin{equation}\label{eq:bsurfsw}
	\esub{x}\dotv\widetilde{\vrm{b}}^\sglt_{{\rm st}}(x_{\rm i}) = \ijot\vrm{k}_{yz}\dotv\vrm{B}^\sglt\, \widetilde{\xi}\;.
\end{equation}
Using \eqref{eq:bstandingwave} this gives
\begin{equation}\label{eq:btilde}
	\widetilde{b}^\sglt 
	= -\signglt\frac{\vrm{k}_{yz}\dotv\vrm{B}^\sglt}{\sin (k_x^{\sglt\mu} \!a_{\rm i}^\sglt)}\, \widetilde{\xi}\;.
\end{equation}

The vacuum magnetic perturbation $\vrm{b}^{\rm v}$ may be found by setting $\mu^\sgt = 0$, in which case \eqref{eq:kxmu} gives $k_x^{\rm v} = \ijot |\vrm{k}_{yz}|$ and \eqref{eq:bstandingwave} becomes \cite{SuppCitePPCF} 
\begin{equation}\label{eq:bvstandingwave}
\begin{split}
	\widetilde{\vrm{b}}^{\rm v}_{\rm st}(x) 
	&= \widetilde{b}^{\rm v}\sinh(|\vrm{k}_{yz}|(x_{\rm w} - x))\,\esub{x} 
	 - \ijot\widetilde{b}^{\rm v}\cosh\left(|\vrm{k}_{yz}|(x_{\rm w} - x)\right)\vrm{k}_{yz}/|\vrm{k}_{yz}| \;,
\end{split}
\end{equation}
where, from \eqref{eq:btilde},
\begin{equation}\label{eq:bvtilde}
	\widetilde{b}^{\rm v} = \frac{\ijot\vrm{k}_{yz}\dotv\vrm{B}^{\rm v}}{\sinh \left(|\vrm{k}_{yz}|(x_{\rm w} - x_{\rm i})\right)}\, \widetilde{\xi}\;.
\end{equation}

\section{Eigenmodes}\label{sec:eigen}

Global eigenvalue relations arise as consistency conditions for perturbed force balance to apply across the 
interface. Taking into account the vanishing of $\grad p^\sglt_0$ and $\grad |\vrm{B}^{\sglt 2}_0|$, \eqref{eq:normsurfvar1} can be written
\begin{equation}\label{eq:eigenvaluecond}
\begin{split}
	(\tau^\slt_0\rho^\slt_1 - \tau^\sgt_0\rho^\sgt_1) + \delta \left(p^\slt_0 - p^\sgt_0\right) 
	&=  \frac{\vrm{B}^\sgt_0\dotv\vrm{B}^\sgt_1 - \vrm{B}^\slt_0\dotv\vrm{B}^\slt_1}{\muSI} 
	+ \delta\frac{B^{\sgt 2}_0 - B^{\slt 2}_0 }{2\muSI}
	\:\:\text{at}\:\: x = x_{\rm i} \;,
\end{split}
\end{equation}
where $\delta\left\{\cdot\right\}$ represents linear fluctuations with the same symmetry as the equilibrium, which are driven adiabatically (due to effectively instantaneous relaxation) by interface fluctuations and are thus as calculated in Sec.~\ref{sec:slabeqpert}. These are associated with oscillations in $\tau_1$ and $\mu_1$, unlike the fluctuations denoted by $\left\{\cdot\right\}_1$, which represent linear fluctuations calculated with $\tau$ and $\mu$ held fixed, defined in Sec.~\ref{sec:slabpert}. The $\delta\left\{\cdot\right\}$ terms pertain only to the $m=0$, $n=0$ modes discussed below, where they can be superposed with the $\left\{\cdot\right\}_1$ terms at linear order. 

In the following we use the notational simplification used in Sec.~\ref{sec:planemag} of suppressing the subscript $0$ on all equilibrium quantities.

\subsection{Purely radial eigenfunctions}\label{sec:mnzero}

First, consider the simple but exceptional case $\vrm{k}_{yz} = 0$ (i.e. $m = n = 0$). As shown in Sec.~\ref{sec:slabeqpert}, in this case $\tau_1(t)$ and $\mu_1(t)$ do not vanish but are easily calculated as equilibrium variations. Also, $\vrm{k}$ is purely radial so the finite-wavelength sound wave perturbations of Sec.~\ref{sec:sound} are functions only of $x$ and $t$ through the factor $\exp(\pm\ijot k^{\sglt\omega}_x - \ijot\omega t)$, where, by \eqref{eq:ksound} with $|\vrm{k}_{yz}|^2$ set to zero,
\begin{equation}\label{eq:ksoundradial}
	 k_x^{\sglt\omega} \equiv \frac{\omega}{C^\sglt_{\rm s}}\;.
\end{equation}

As in Sec.~\ref{sec:planemag} the subscript 0 on equilibrium quantities is implicit throughout this section. 
Using \eqref{eq:pvar} and \eqref{eq:rhostandingwave} we find the LHS of \eqref{eq:eigenvaluecond} \begin{equation}\label{eq:radialeigvalLHS} 
\begin{split}
	&\tau^\slt\rho^\slt_1 - \tau^\sgt\rho^\sgt_1  + \delta (p^\slt - p^\sgt) 
	= -\Re\!\sum_\sglt \rho^\sglt\left[\frac{\omega^2\cot(k^{\sglt\omega}_x a_{\rm i}^\sglt)}{k^{\sglt\omega}}
	+ \frac{\gamma \tau^\sglt}{a_{\rm i}^\sglt} \right] \widetilde{\xi}\,\e^{-\ijot\omega t} \\
\end{split}
\end{equation}
There is no contribution to force balance from a vacuum or pressureless plasma in the outer region, $\Omega^\sgt$, so in this case contributions from $\Omega^\sgt$ must be deleted from \eqref{eq:radialeigvalLHS}. The issue of interpreting $k_x^{\sgt\omega}$ in the limit $C^\sgt_{\rm s} \to 0$ thus does not arise.

In MRxMHD both relaxed and vacuum magnetic fields instantaneously adjust to the radial oscillations, so, as far as the magnetic field is concerned, the perturbed states are equivalent to the varied equilibrium states of Sec.~\ref{sec:slabeqm}. 
Setting $\vrm{B}^\sglt_1 = 0$ (as the magnetic response is fully accounted for by $\delta\vrm{B}_0$ when $\vrm{k}_{yz} = 0$) 
and using \eqref{eq:B0var} we find the RHS of \eqref{eq:eigenvaluecond}, \cite{SuppCitePPCF}
\begin{equation}\label{eq:radialeigvalRHS} 
\begin{split}
	\delta\frac{B^{\sgt 2} - B^{\slt 2} }{2\muSI}
	&=\Re\!\sum_\sglt\frac{\rho^\sglt v^{\sglt 2}_{\rm A}}{a_{\rm i}^\sglt} \, \widetilde{\xi}\,\e^{-\ijot\omega t} \;,
\end{split}
\end{equation}
where we have eliminated the $B^\sglt$ in favour of the Alfv\'en speeds in the two regions [cf. \eqref{eq:betaalt} ff.]. 

For the two sides of \eqref{eq:eigenvaluecond} we use \eqref{eq:radialeigvalLHS} (with $\omega$ eliminated in favour of $k^{\sglt\omega}_x$ \cite{SuppCitePPCF}) 
and \eqref{eq:radialeigvalRHS} to find the eigenvalue condition for radial oscillations
\begin{equation}\label{eq:radialeigvalcond}
	-\sum_\sglt \left(\frac{\rho^\sglt C^{\sglt 2}_{\rm s}}{a_{\rm i}^\sglt}\right)
		k_x^{\sglt\omega}a_{\rm i}^\sglt\cot(k^{\sglt\omega}_x a_{\rm i}^\sglt) \\
	= \sum_\sglt\frac{\rho^\sglt \left(v^{\sglt 2}_{\rm A} + \gamma C^{\sglt 2}_{\rm s}\right)}{a_{\rm i}^\sglt} \;.
\end{equation} 

\begin{figure}[htbp]
   \centering
		\includegraphics[width = 0.7\textwidth]{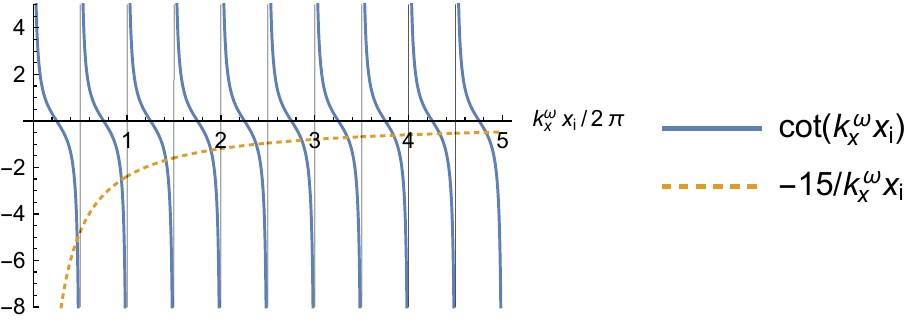} 
\caption{Illustrating the large-$\mathcal{V}$ asymptotics of the spectrum of purely radial oscillations using the case $\mathcal{V} = 15$. The intersections of the two curves give the solutions of \eqref{eq:radialeigvalcondsimple}.}
\label{fig:radEigVals}
\end{figure}

As mentioned above, in the case of a vacuum or pressureless plasma in the outer region, sonic contributions from $\Omega^\sgt$ must be deleted from \eqref{eq:radialeigvalcond}, which can be done formally by setting $C_{\rm s}^{\sgt 2}$ to zero. However, as $\rho^\sgt v^{\sgt 2}_{\rm A} = B^{\sglt 2}/\muSI$ is finite as $\rho^\sgt \to 0$, the vacuum magnetic field contribution remains well defined. In this case the eigenvalue equation is particularly simple and can be written in the form
\begin{equation}\label{eq:radialeigvalcondsimple}
	k_x^{\omega}x_{\rm i}\cot(k_x^{\omega} x_{\rm i}) = -\mathcal{V} \;,
\end{equation}
where the positive dimensionless parameter $\mathcal{V}$ is defined by \cite{SuppCitePPCF} 
\begin{equation}\label{eq:radialW}
\begin{split}
	 \mathcal{V} &\equiv  \gamma + \left(\frac{v^{\slt}_{\rm A}}{C^{\slt}_{\rm s}}\right)^2 \left[1 +
	 				\frac{x_{\rm i}}{(x_{\rm w} - x_{\rm i})}\frac{B^{\sgt 2}}{B^{\slt 2}} \right] \;.
\end{split}
\end{equation}

Figure~\ref{fig:radEigVals} illustrates a graphical method for solving \eqref{eq:radialeigvalcondsimple}, by finding the intersections of the graphs of $\cot(k_x^{\omega} x_{\rm i})$ and $-\mathcal{V}/(k_x^{\omega} x_{\rm i})$. Once solutions $k_x^{\omega}$ are known the spectrum of eigenvalues $\omega$ is obtained immediately from the dispersion relation \eqref{eq:sounddisprel}.

By \eqref{eq:betaalt} $(v^{\slt}_{\rm A}/C^{\slt}_{\rm s})^2 = 2/\beta^\slt$. Thus, in a low-$\beta$ plasma [and/or if $x_{\rm w} - x_{\rm i} \to 0$], $\mathcal{V}$ is a large number. In this case the eigenvalues are given approximately by the ``waves on a string'' spectrum (Fig.~\ref{fig:radEigVals}),
\begin{equation}\label{eq:radialeigvals}
	 k_x^{\omega}x_{\rm i} \approx \pi l \;, \:\: l = 1,2,\ldots \;,
\end{equation}
where $l$ is the radial mode number. In terms of frequency [cf. \eqref{eq:sounddisprel}] $\omega = \omega_l \approx \pi l C_{\rm s}/x_{\rm i}$, a sequence consisting of the fundamental, $\omega_1$, and its harmonics. Like an organ pipe open at one end, the higher-order modes are not exact harmonics --- as can be seen from Fig.~\ref{fig:radEigVals}, $\omega_l \approx \pi (l - 1/2)C_{\rm s}/x_{\rm i}$ as $l \to \infty$. 

Note that in MRxMHD the high-frequency fast magnetosonic mode within relaxation regions is eliminated because of the decoupling of velocity and magnetic field perturbations. This should allow time  steps longer than those that can be used in ideal-MHD numerical simulations using explicit methods \cite[e.g. \S 2.2]{Jardin_10}.

To determine stability we see from \eqref{eq:instability} that we need to seek roots of \eqref{eq:radialeigvalcond} on the imaginary axis in the complex $k_x^{\omega}$ plane, where the LHS of \eqref{eq:radialeigvalcond} is $-\sum_\sglt (\rho^\sglt C^{\sglt 2}_{\rm s}/a_{\rm i}^\sglt)|k_x^{\sglt\omega}a_{\rm i}^\sglt| \coth(|k_x^{\sglt\omega} a_{\rm i}^\sglt|)$. This is negative for all $|k_x^{\omega} x_{\rm i}|$, while the RHS is positive. Thus there are no unstable purely radial ($\vrm{k}_{yz} = 0$) modes. 

\subsection{Surface wave eigenvalue problem}\label{sec:mngen}

When $\vrm{k}_{yz} \neq 0$ the only change required to the expression in \eqref{eq:radialeigvalLHS} for the LHS of \eqref{eq:eigenvaluecond} is to delete the term in $\gamma$ and to use the full expression \eqref{eq:ksound} for $k^{\sglt\omega}_x$ rather than the simplified expression in \eqref{eq:ksoundradial}.

However the calculation of the RHS side of \eqref{eq:eigenvaluecond} is quite different --- the $\delta$ term vanishes whereas $\vrm{B}_1 \equiv \vrm{b}$ does not, being given by \eqref{eq:bstandingwave}. 
To satisfy linearized force balance we equate the RHS of \eqref{eq:radialeigvalLHS} to $(\vrm{B}^\sgt\dotv\vrm{b}^\sgt - \vrm{B}^\slt\dotv\vrm{b}^\slt)/\muSI$ and multiply both sides by the non-dimensionalizing factor $a^0\muSI/B^{02} \equiv a^0/\rho^0v^{0\,2}_{\rm A}$, where $B^0$ is any convenient reference magnetic field strength, with corresponding Alfv\'en speed $v^{0}_{\rm A} \equiv B^0/\sqrt{\muSI \rho^0}$ [cf. \eqref{eq:betaalt} ff.], and the constants $a^0$ and $\rho^0$ are any convenient reference length and mass density, respectively. 

This gives the eigenvalue-like equation
\begin{equation}\label{eq:surfeigvalue}
	\mathcal{K}(\lambda)\,\lambda = \mathcal{W} \;,
\end{equation}
where the factors on the LHS of \eqref{eq:surfeigvalue} are the \emph{dimensionless eigenvalue} 
\begin{subequations}
\begin{align}
	\lambda &\equiv \left(\frac{a^0 \omega}{v^{0}_{\rm A}}\right)^2 \;, \label{eq:lambdadef} \\
	& = \left(\frac{C_{\rm s}^\sglt}{v^{0}_{\rm A}}\right)^2 [(|\vrm{k}_{yz}|a^0)^2 + (k_x^{\sglt\omega}a^0)^2]
	\label{eq:lambdaalt} \;,
\end{align}
\end{subequations}
and the \emph{dimensionless normalization factor}
\begin{subequations}
\begin{align}
	\mathcal{K}
	&\equiv -\sum_\sglt \frac{\rho^\sglt a_{\rm i}^\sglt}{\rho^0 a^0}
	\frac{\cot(k^{\sglt\omega}_x a_{\rm i}^\sglt)}{k_x^{\sglt\omega}a_{\rm i}^\sglt} \label{eq:surfwKdefRe}\\
	&= \sum_\sglt \frac{\rho^\sglt a_{\rm i}^\sglt}{\rho^0 a^0}
	\frac{\coth(|k^{\sglt\omega}_x| a_{\rm i}^\sglt)}{|k_x^{\sglt\omega}|a_{\rm i}^\sglt} \label{eq:surfwKdefIm}
	\;, \: k^{\sglt\omega}_x =\ijot|k^{\sglt\omega}_x| \;.
\end{align}
\end{subequations}
Equation~(\ref{eq:lambdaalt}) makes explicit the relation between the radial sound wavenumbers $k_x^{\sglt\omega}$ and $\lambda$ using \eqref{eq:sounddisprel}. By \eqref{eq:betaalt} the factors $(C_{\rm s}^\sglt/v^{0}_{\rm A})^2$ can also be written as $\beta^{0\,\sglt}/2$, where $\beta^{0\,\sglt} \equiv 2\muSI p^\sglt/B^{0\,2}$.

Note that we derived the form in \eqref{eq:surfwKdefIm} using the identity $\ijot u \cot \ijot u = u \coth u$, which will also be useful in the expression for $\mathcal{W}$ given below if $k_x^{\sglt\mu}$ is imaginary. From \eqref{eq:kxmu} this occurs when $|\mu^\sglt| < |\vrm{k}_{yz}|$.

Being proportional to $\omega^2$, the dimensionless eigenvalue $\lambda$ is \emph{always real}: $\lambda \geq 0$ for stable modes and $\lambda<0$ for unstable modes. It is thus more convenient to use in stability studies than $\omega$. The expression for $\mathcal{K}$ in \eqref{eq:surfwKdefRe} is valid for all $\lambda$, but most appropriate to cases where both $k^{\sglt\omega}_x$ are real, i.e. for $\lambda > (|\vrm{k}_{yz}|a^0\max_\sglt C_{\rm s}^\sglt/v^{0}_{\rm A})^2$, by \eqref{eq:ksound} and \eqref{eq:lambdadef}.

On the other hand, the manifestly positive form in \eqref{eq:surfwKdefIm} is specialized to cases where both $k^{\sglt\omega}_x$ are imaginary. This includes \emph{all unstable modes}, $\lambda < 0$, and a range of stable modes close to the instability threshold, $0 \leq \lambda < (|\vrm{k}_{yz}|a^0\min_\sglt C_{\rm s}^\sglt/v^{0}_{\rm A})^2$. When $k^{\sglt\omega}_x$ is imaginary, \cite{SuppCitePPCF} 
\begin{equation}\label{eq:ksoundIm} 
	|k_x^{\sglt\omega}|
	= \frac{1}{a^0}\left[(|\vrm{k}_{yz}| a^0)^2 - \frac{2\lambda}{\beta^{0\,\sglt}} \right]^{1/2} \;,
\end{equation}
from \eqref{eq:lambdaalt}.
As $\mathcal{K}$ is positive definite in this case it can be regarded as a frequency-dependent effective mass reminiscent of the artificial constant surface mass normalization used in our previous stability studies \cite{Hole_Hudson_Dewar_07,Mills_Hole_Dewar_09, Hole_Mills_Hudson_Dewar_09}. In fact, close enough to marginal stability (see Sec.~\ref{sec:marginal}) we may approximate $\mathcal{K}$ by its value at $\lambda = 0$ to give the Rayleigh--Ritz-like approximation 
\begin{equation}\label{eq:incompressible}
	\lambda \approx \frac{\mathcal{W}}{\mathcal{K}(0)} \;.
\end{equation}
 As $\divv\vrm{v} = 0$  at $\lambda = 0$ we shall term this the \emph{incompressible approximation}.\footnote{From \eqref{eq:momentumeq1F}, $\vrm{k}^{{\rm s}\pm}\dotv\widetilde{\vrm{v}}^{\pm} \propto |\vrm{k}_{yz}|^2 - |k_x^{\omega}|^2$, so $\divv\vrm{v}$ vanishes when $\lambda$ vanishes.}

The RHS of \eqref{eq:surfeigvalue} is the \emph{dimensionless energy}, defined \cite{SuppCitePPCF} as 
\begin{equation}\label{eq:surfwWdef}
	\mathcal{W} = 
	\sum_\sglt a^0 k_x^{\sglt\mu}\cot (k_x^{\sglt\mu} a_{\rm i}^\sglt) [F^\sglt_{m,n}(x_{\rm i})]^2
	+\jump{ a^0\!\mu\, F_{m,n}G_{m,n} }_\slgt
	\;,
\end{equation}
where
\begin{align}
	F^\sglt_{m,n}(x) &\equiv \frac{\vrm{k}_{yz}\dotv\vrm{B}^\sglt(x)}{|\vrm{k}_{yz}|B^0}
	= \frac{B^\sglt}{B^0}\,\frac{m\sin(\Theta^\sglt + \mu^\sglt x) - \epsilon_{\rm a} n\cos(\Theta^\sglt + \mu^\sglt x)}{(m^2 + \epsilon_{\rm a}^2 n^2)^{1/2}} \;, \label{eq:Fdef}\\
	G^\sglt_{m,n}(x) &\equiv \frac{\esub{x}\dotv\vrm{k}_{yz}\cross\vrm{B}^\sglt(x)}{|\vrm{k}_{yz}|B^0} 
	= \frac{B^\sglt}{B^0}\,\frac{\epsilon_{\rm a} n\sin(\Theta^\sglt + \mu^\sglt x) + m\cos(\Theta^\sglt + \mu^\sglt x)}{(m^2 + \epsilon_{\rm a}^2 n^2)^{1/2}} \;. \label{eq:Gdef}
\end{align}

In \eqref{eq:surfwWdef}, $\jump{\cdot}_\slgt$ denotes the jump from $\Omega^\slt$ to $\Omega^\sgt$ at $x_{\rm i}$. The jump term vanishes if $\vrm{B}$ and $\mu$ are continuous at the interface (in which case there is no equilibrium current sheet, but the interface is still assumed to act as an ideal-MHD barrier to relaxation).

\section{Vacuum case}\label{sec:vac}

Provided there is no linear correction to $\mu$, which is ensured here because $\vrm{k}_{yz} \neq 0$, the case [designated by superscript (v)] where there is a \emph{vacuum} in $\Omega^\sgt$ can be treated as if the vacuum were a currentless plasma. That is, by taking the limit $\mu^\sgt \to 0$, a case where $k_x^{\sgt\mu}$ is imaginary. Also, to make the region pressureless, we take $\rho^\sgt \to 0$. (However, without physical consequence we can keep the specific temperature $\tau^\sgt$ finite in order to avoid any complications from vanishing $C_{\rm s}^\sgt$.) 

With plasma confined to $\Omega^\slt$ it is natural to use the parameters of this region as reference parameters, i.e. to set $a^0 = a_{\rm i}^\slt \equiv x_{\rm i}$,  $\rho^0 = \rho^\slt$ and $B^0 = B^\slt$ (so $v^{0}_{\rm A} = B^\slt/\sqrt{\muSI \rho^\slt}$). 
Then the eigenvalue equation \eqref{eq:surfeigvalue} simplifies to $\mathcal{K}^{(\rm v)}\lambda = \mathcal{W}^{(\rm v)}$, where
\begin{equation}\label{eq:surfwWdefvac}
\begin{split}
\mathcal{W}^{(\rm v)} &\equiv
	\frac{[\vrm{k}_{yz}\dotv\vrm{B}(x_{\rm i})]^2\,k_x^{\mu}x_{\rm i} \cot k_x^{\mu} x_{\rm i} }{|\vrm{k}_{yz}|^2 B^2} 
	 -\frac{\mu x_{\rm i} \vrm{k}_{yz}\dotv\vrm{B}(x_{\rm i}) \,\esub{x}\dotv\vrm{k}_{yz}\cross\vrm{B}(x_{\rm i}) }{|\vrm{k}_{yz}|^2 B^2} \\
	&\quad\quad\quad\quad\quad + \frac{B_{\rm v}^2}{B^2}\frac{ |\vrm{k}_{yz}| x_{\rm i}(\vrm{k}_{yz}\dotv\vrm{B}_{\rm v})^2 \coth |\vrm{k}_{yz}|(x_{\rm w} - x_{\rm i})}{|\vrm{k}_{yz}|^2 B_{\rm v}^2}
	\;, \\
	&= [F_{m,n}(x_{\rm i})]^2\,k_x^{\mu}x_{\rm i} \cot k_x^{\mu} x_{\rm i}  
	 - \mu x_{\rm i} F_{m,n}(x_{\rm i}) G_{m,n}(x_{\rm i}) \\
	& 
	\quad +(1+\beta)|\vrm{k}_{yz}| x_{\rm i}[F^{\rm v}_{m,n}]^2 \coth |\vrm{k}_{yz}|(x_{\rm w} - x_{\rm i})
	\;. 
\end{split}
\end{equation}
The superscripts ${}^\sglt$ have been dropped because all plasma parameters are defined only in $\Omega^\slt$ and the vacuum in $\Omega^\sgt$ is indicated by subscript v. We have eliminated the vacuum magnetic field strength in terms of the plasma $\beta$ using \eqref{eq:normsurfvar0}: $B_{\rm v} = B(1 + \beta)^{1/2}$. 

Using \eqref{eq:lambdaalt} to eliminate $\lambda$, the LHS of the eigenvalue equation can be put in a form reminiscent of  \eqref{eq:radialeigvalcondsimple},
\begin{subequations}
\begin{align}
	\mathcal{K}^{(\rm v)}\lambda
	&\equiv -\frac{\beta}{2}\,\frac{(|\vrm{k}_{yz}| x_{\rm i})^2 + (k_x^{\omega}x_{\rm i})^2}{k_x^{\omega} x_{\rm i}}\cot(k_x^{\omega} x_{\rm i}) \label{eq:surfwKvdefRe}\\
	&= \frac{\beta}{2}\,\frac{(|\vrm{k}_{yz}| x_{\rm i})^2 - |k_x^{\omega} x_{\rm i}|^2}{|k_x^{\omega} x_{\rm i}|}
	\coth(|k_x^{\omega} x_{\rm i}|) \label{eq:surfwKvdefIm}
	\;, \: k^{\omega}_x =\ijot|k^{\omega}_x| 
	 \;,
\end{align}
\end{subequations}
so the numerical root-finding required to find eigenvalues can be done using $k_x^{\omega} x_{\rm i}$ as independent variable rather than $\lambda$, which can then be determined from \eqref{eq:lambdaalt}.

As $\mathcal{W}$ is not positive definite the spectrum contains unstable modes in general: the threshold between stability and instability can be found by finding where $\mathcal{W}$ changes sign, causing $\lambda$ also to change sign as $\mathcal{K}$ is positive. Unlike the kinetic energy in ideal MHD, $\mathcal{K}$ is an analytic function of $\lambda$ through the marginal stability point $\lambda = 0$, allowing instability thresholds to be determined by simple interpolation methods in a scan of equilibrium parameters (a feature similar to the PEST 2 code \cite[Fig. 4]{Grimm_Dewar_Manickam_83}, which also has a kinetic energy depending only on displacements normal to magnetic surfaces). 

From \eqref{eq:surfwWdef} we see the potentially negative factors in $\mathcal{W}$ are the  second (jump) term and the $\cot(k_x^{\sglt\mu} a_{\rm i}^\sglt)$ factors in the first term. A simple example of the latter is given in Sec.~\ref{sec:Tearing} below.

\subsection{Spectrum in case of radially propagating sound}\label{sec:propagating}

\begin{figure}[htbp]
   \centering
		\includegraphics[width = 0.9\textwidth]{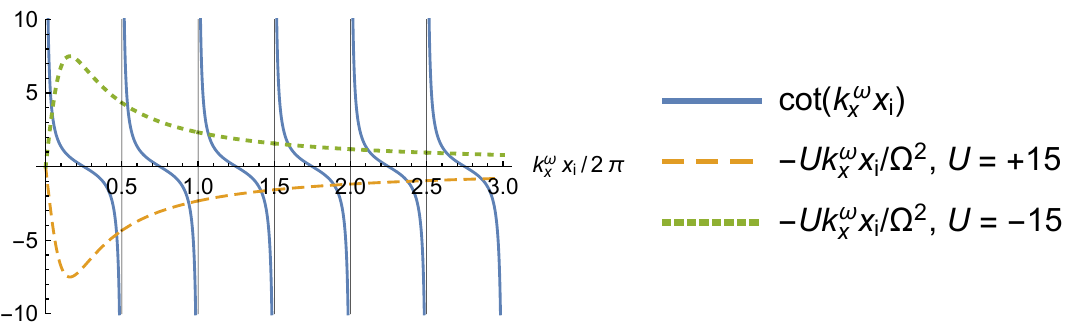} 
\caption{Case of radially propagating sound, $k_x^{\omega}$ real: there is an infinite number of stable ($\lambda > 0$) surface modes whether $\mathcal{W}^{(\rm v)}$ is positive or negative, illustrated for $\mathcal{U} = + 15$ (long-dashed orange curve), and $\mathcal{U} = - 15$ (short-dashed green curve). We have taken $|\vrm{k}_{yz}|x_{\rm i} = 1$, so that $\Omega^2 = (k_x^{\omega} x_{\rm i})^2 + 1$. The intersections of the graph of $\cot (k_x^{\omega} x_{\rm i})$ (solid blue curve) with the other two curves give solutions corresponding to  \eqref{eq:surfwKvdefRe}. (Colour online.)}
\label{fig:surfwRekxs}
\end{figure}

When $k_x^{\omega} x_{\rm i}$ is real the spectrum of eigenvalues $\omega$, or equivalently $\lambda$, is determined by finding standing waves in the $x$-direction. To get an overview of this $k_x^{\omega} \in \mathbb{R}$ subset of modes [which are all stable as $\lambda > (\beta/2) |\vrm{k}_{yz}|^2 x_{\rm i}^2 > 0$] consider solutions of $\mathcal{K}^{(\rm v)}\lambda = \mathcal{W}^{(\rm v)}$ on the real-$k_x^{\omega}$ axis, corresponding to the form of $\mathcal{K}^{(\rm v)}\lambda$ in \eqref{eq:surfwKvdefRe}. As seen in Fig.~\ref{fig:surfwRekxs}, solutions may be determined by finding points where the graphs of $\cot(k_x^{\omega} x_{\rm i})$ and $-\mathcal{U}\, k_x^{\omega} x_{\rm i}/\Omega^2$ intersect, where $\mathcal{U} \equiv 2\mathcal{W}^{(\rm v)}/\beta$ and $\Omega^2 \equiv (|\vrm{k}_{yz}|^2+k_x^{\omega 2})x_{\rm i}^2 > 0$. It is seen that these positive MRxMHD eigenvalues form an infinite point spectrum rather than the continuum found in ideal MHD. (However, the MRxMHD stable spectrum does become dense as $\beta \to 0$ and/or if the number of interfaces approaches infinity.) Note that, when $\mathcal{U} \ll -1$, the lowest root in the propagating sound subset is close to $k_x^{\omega} x_{\rm i} = 0$ (i.e. $\lambda = \beta |\vrm{k}_{yz}|^2 x_{\rm i}^2/2$), while in the case $\mathcal{U} \gg 1$, the lowest root is close to $k_x^{\omega} x_{\rm i} = \pi$ [i.e. $\lambda = (\beta/2) (|\vrm{k}_{yz}|^2 x_{\rm i}^2 + \pi^2)$].

\subsection{Spectrum in case of radially evanescent sound}\label{sec:evanescent}

\begin{figure}[htbp]
   \centering
		\includegraphics[width = 0.9\textwidth]{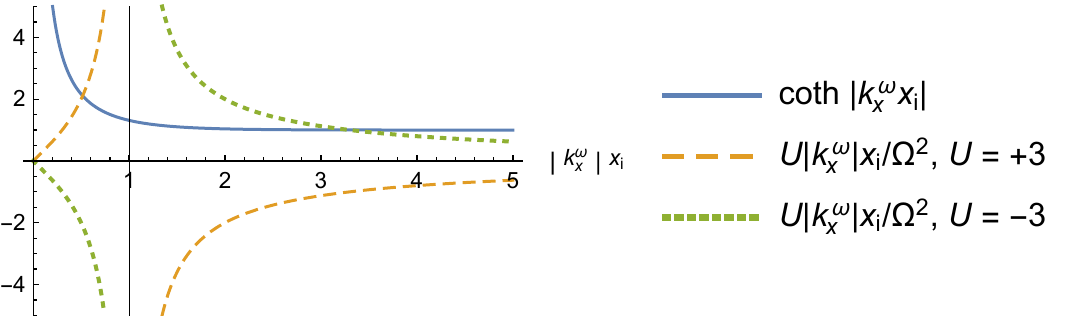} 
\caption{Case of radially evanescent sound, $k_x^{\omega}$ imaginary: the intersections of the graph of $\coth |k_x^{\omega} x_{\rm i}|$ (solid blue curve) with the other two curves give eigenmode solutions corresponding to \eqref{eq:surfwKvdefIm}. The vertical \emph{marginal stability} line at $|k_x^{\omega}| = |\vrm{k}_{yz}|$,  separates stable modes ($\lambda > 0$, $|k_x^{\omega}| < |\vrm{k}_{yz}|$) from unstable modes ($\lambda < 0$, $|k_x^{\omega}| > |\vrm{k}_{yz}|$). For given $\mathcal{W}^{(\rm v)}$ there is only one root --- stable for positive $\mathcal{W}^{(\rm v)}$, unstable for negative.  Illustrated for $\mathcal{U} = + 3$ (long-dashed orange curve) and $\mathcal{U} = - 3$ (short-dashed green curve).  (Colour online.)}
\label{fig:surfwImkxs}
\end{figure}

To get an overview of the nature of modes for which $(k_x^{\omega} x_{\rm i})^2$ is negative, $k_x^{\omega} x_{\rm i} \in \mathbb{I}$ (when $\lambda <  \beta |\vrm{k}_{yz}|^2 x_{\rm i}^2/2$) we use the same approach as above, but with the form of $\mathcal{K}^{(\rm v)}\lambda$ in \eqref{eq:surfwKvdefIm}. As seen in Fig.~\ref{fig:surfwImkxs}, solutions may be determined by finding points where the graphs of $\coth(|k_x^{\omega} x_{\rm i}|)$ and $\mathcal{U} |k_x^{\omega} x_{\rm i}|/\Omega^2$ intersect, where $\Omega^2 = (|\vrm{k}_{yz}|^2-|k_x^{\omega}|^2)x_{\rm i}^2$ is now less than $|\vrm{k}_{yz}|^2 x_{\rm i}^2$ and passes through zero (the marginal stability point, $\lambda = 0$) when $|k_x^{\omega}| = |\vrm{k}_{yz}|$. In contrast to the propagating case it is seen that there is only one root in the evanescent sound case, which is stable if $\mathcal{W}^{(\rm v)} > 0$, but is unstable if $\mathcal{W}^{(\rm v)} < 0$, consistently with the interpretation of $\mathcal{W}^{(\rm v)}$ as a non-dimensionalized second variation of the MHD energy \cite{Hole_Hudson_Dewar_07,Mills_Hole_Dewar_09, Hole_Mills_Hudson_Dewar_09} (this connection will be pursued further elsewhere).

In the strongly stable case $\mathcal{U} \gg 1$ there is a root close to $k_x^{\omega} x_{\rm i} = 0$, but, in the strongly unstable case $\mathcal{U} \ll -1$, $|k_x^{\omega}| x_{\rm i}$ (and hence $-\lambda$) are large, as shown below.

\subsection{Asymptotics}\label{sec:asymp}

\subsubsection{Close-to-marginal modes:}\label{sec:marginal}

As seen in Fig.~\ref{fig:surfwImkxs}, roots close to marginal stability occur close to the vertical asymptote at $|k_x^{\omega}x_{\rm i}| = |\vrm{k}_{yz}x_{\rm i}|$. In this case we expand $\coth(|k_x^{\omega} x_{\rm i}|)$ about $|\vrm{k}_{yz}x_{\rm i}|$ in \eqref{eq:surfwKvdefIm}, $\mathcal{K}^{(\rm v)}\lambda = \mathcal{W}^{(\rm v)} = (\beta/2)\mathcal{U}$ 
then giving $|k_x^{\omega} x_{\rm i}| \sim |\vrm{k}_{yz}x_{\rm i}| - \tanh(|\vrm{k}_{yz}x_{\rm i}|)\mathcal{U}/2 + O(\mathcal{U}^2)$.
From \eqref{eq:lambdaalt} the lowest eigenvalue is found to be
\begin{equation}\label{eq:lambdamarginal}
\begin{split}
	\lambda_1 &\sim |\vrm{k}_{yz}x_{\rm i}|\tanh (|\vrm{k}_{yz}x_{\rm i}|)\mathcal{W}^{\rm (v)} \\
	&\quad -\frac{\tanh(|\vrm{k}_{yz}x_{\rm i}|)}{\beta}
		\left[|\vrm{k}_{yz}x_{\rm i}|\sech^2(|\vrm{k}_{yz}x_{\rm i}|) + \tanh(|\vrm{k}_{yz}x_{\rm i}|)\right]\mathcal{W}^{\rm (v)2}
	+ O(\mathcal{W}^{\rm (v)3})
\end{split}
\end{equation}
as $\mathcal{W}^{\rm (v)} \to 0$.

The striking thing about this result, at leading order, is that the lowest eigenvalue is independent of $\beta$. This can also be seen by substituting the $\lambda = 0$ value $|k_x^{\omega}x_{\rm i}| = |\vrm{k}_{yz}x_{\rm i}|$ [see \eqref{eq:ksoundIm}] into \eqref{eq:surfwKdefIm}, giving $\mathcal{K}^{\rm (v)}(0) = \coth(|\vrm{k}_{yz}x_{\rm i}|)/|\vrm{k}_{yz}x_{\rm i}|$. Thus the incompressible approximation \eqref{eq:incompressible} gives $\lambda \approx  |\vrm{k}_{yz}x_{\rm i}|\tanh (|\vrm{k}_{yz}x_{\rm i}|)\mathcal{W}^{\rm (v)}$. This shows that, near marginal stability, \emph{the evanescent sound waves couple plasma inertia to the interface independent of the value of the plasma pressure}. (However, the range of $\mathcal{W}^{\rm (v)}$ over which this slow, incompressible approximation is appropriate shrinks to zero as $\beta \to 0$.)

Thus our previous approach \cite{Hole_Hudson_Dewar_07,Mills_Hole_Dewar_09, Hole_Mills_Hudson_Dewar_09} of assigning an artificial, constant mass loading to the interface gives identical stability boundaries to those found from the present dynamical MRxMHD formulation.

\subsubsection{Strongly unstable modes:}\label{sec:unstablelargenegU}

For large $|k_x^{\omega} x_{\rm i}|$ in Fig.~\ref{fig:surfwImkxs}, $\coth|k_x^{\omega} x_{\rm i}| \approx 1 + 2 \exp(-2|k_x^{\omega} x_{\rm i}|)$. However, the exponential term is small to all orders in $1/|k_x^{\omega} x_{\rm i}|$, so may be dropped to find an asymptotic solution of $\mathcal{K}^{(\rm v)}\lambda = (\beta/2)\mathcal{U}$ 
for the strongly unstable case $\mathcal{W}^{\rm (v)} \to -\infty$, $k_x^{\omega} \in \mathbb{I}$,
$|k_x^{\omega}| x_{\rm i} \sim |\mathcal{U}| + (|\vrm{k}_{yz}|x_{\rm i})^2|\mathcal{U}|^{-1} + O\left(|\mathcal{U}|^{-3}\right)$ \cite{SuppCitePPCF}. 
Thus, from \eqref{eq:lambdaalt}, the lowest eigenvalue as $\mathcal{W}^{\rm (v)} \to -\infty$ is
\begin{equation}\label{lambdalargekxs}
	\lambda_1 \sim
	-\frac{2}{\beta}|\mathcal{W}^{\rm (v)}|^2 - \frac{\beta}{2}|\vrm{k}_{yz}x_{\rm i}|^2 + O(|\mathcal{W}^{\rm (v)}|^{-2}) \;.
\end{equation}

\subsubsection{Stable modes with large $|\mathcal{U}|$:}\label{sec:stablelargeU}

As seen qualitatively in Figs.~\ref{fig:surfwRekxs} and \ref{fig:surfwImkxs}, in both the propagating and evanescent sound cases there may (depending on the sign of $\mathcal{W}^{(\rm v)}$) be a root where $k_x^{\omega} x_{\rm i}$ is close to zero. In these cases we can expand the eigenvalue equation $\mathcal{K}^{(\rm v)}\lambda = (\beta/2)\mathcal{U}$ 
in powers of $(k_x^{\omega} x_{\rm i})^2$ and solve for $(k_x^{\omega} x_{\rm i})^2$, giving
$(k_x^{\omega} x_{\rm i})^2 \sim -(|\vrm{k}_{yz}|x_{\rm i})^2 \mathcal{U}^{-1} + O\left(|\mathcal{U}|^{-2}\right)$.
Then
\begin{equation}\label{eq:lambdasmallkxs}
	\lambda \sim \frac{\beta|\vrm{k}_{yz}|^2 x_{\rm i}^2}{2}\left(1 - \frac{\beta}{2\mathcal{W}^{\rm (v)}}\right)
	+ O(\mathcal{W}^{\rm (v)-2}) \;,
\end{equation}
for the lowest stable mode in the limit $|\mathcal{W}^{(\rm v)}| \to \infty$ (i.e. the lowest mode $\lambda_1$ when $\mathcal{W}^{(\rm v)} \to +\infty$, $k_x^{\omega} \in \mathbb{I}$ and the second lowest mode $\lambda_2$ when $\mathcal{W}^{(\rm v)} \to -\infty$, $k_x^{\omega} \in \mathbb{R}$).

\subsection{$\mu$ scan}\label{sec:muscan} 


\begin{figure}[htbp]
   \centering
		\includegraphics[width = 0.9\textwidth]{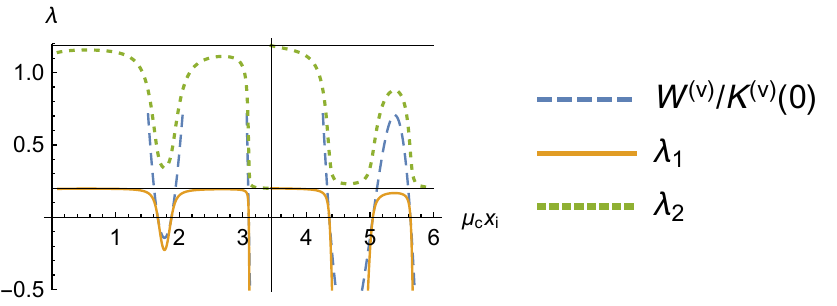} 
\caption{Plot of $\lambda_1$ (solid orange curve), $\lambda_2$ (short-dashed green curve), and the incompressible approximation to $\lambda_1$ (long-dashed blue curve) \emph{vs.} $\mu_{\rm c}$. The parameter values are $R = x_{\rm i} = a$, $x_{\rm w}/x_{\rm i} = 1.2$, $m = 1$, $n = -1$, $\beta = 0.2$.
(Colour online.)}
\label{fig:lambdamuscan}
\end{figure}

To illustrate these various limits we generate a family of (rather artificial) equilibria in a similar way to those depicted in Sec.~\ref{sec:slabeqm}, but choosing $q^\slt(0)$ at the core-slab interface to be the zero-radius core value $q_{\rm c}(0)$ given in \eqref{eq:cylqs} so that $q^\slt(0) = 2/R \mu_{\rm c}$. Then \eqref{eq:qglt} can be solved \cite[Appendix B]{SuppCitePPCF} to find $\Theta^\slt = \tan^{-1}(a\mu_{\rm c}/2)$. We also choose $\mu^\slt = \mu_{\rm c}$. Figure~D1 of \cite{SuppCitePPCF} plots $\mathcal{W}^{\rm (v)}$ \emph{vs.} $\mu_{\rm c}x_{\rm i}$, showing several stable ($\mathcal{W}^{\rm (v)} > 0$) and unstable ($\mathcal{W}^{\rm (v)} < 0$) ranges of $\mu_{\rm c}$. 

Figure~\ref{fig:lambdamuscan} shows plots of eigenvalues $\lambda$ \emph{vs.} $\mu_{\rm c}$: the lowest eigenvalue $\lambda_1$ (negative when the system is unstable) and the second lowest eigenvalue $\lambda_2$. These eigenvalues were found by solving the eigenvalue equation $\mathcal{K}^{(\rm v)}\lambda = \mathcal{W}^{(\rm v)}$ numerically. The incompressible approximation to $\lambda_1$, $\mathcal{W}^{\rm (v)}/\mathcal{K}^{\rm (v)}(0)$ [\eqref{eq:lambdamarginal} ff.], is also plotted, showing good agreement with the numerical solution when the system is close to marginal stability, $\lambda_1 \approx 0$. 

The lower horizontal line at height $ (\beta/2) |\vrm{k}_{yz}|^2 x_{\rm i}^2 $ divides the range of radially propagating sonic waves (above) from the radially evanescent lowest eigenmode range (below), and the upper line at height $(\beta/2) (|\vrm{k}_{yz}|^2 x_{\rm i}^2 + \pi^2)$ separates the range of possible $\lambda_2$ values from the $\lambda_3$ range. The vertical line indicates the position of the first Beltrami eigenvalue \eqref{eq:BeltramiEigval}, $\mu_1^{1,-1} = 3.445/x_{\rm i}$, if the wall were at $x_{\rm i}$. 

\section{Free and fixed-boundary tearing instabilities}\label{sec:Tearing}

Consider first the case where there is no vacuum region and the unperturbed inner and outer regions are simply subregions of the same Taylor-relaxed plasma, confined by rigid plane boundaries at $x=0$ and $x = x_{\rm w}$ and partitioned by a thin ideal interface at arbitrary $x_{\rm i}$. Then $\vrm{B}^{\sglt}(x)$ are analytic continuations of each other, as depicted by the dashed curve in Fig.~\ref{fig:qprofTok}, and the jump term in \eqref{eq:surfwWdef} vanishes. 

In addition to the interface at $x = x_{\rm i}$ and the outer boundary at $x = x_{\rm w}$, an important location in the plasma is, for given $\vrm{k}^{m,n}_{yz}$ [see \eqref{eq:kmn}], the \emph{mode rational surface} at $x = x^{m,n}_{\rm r}$, defined as the solution of the ideal-MHD resonance condition at marginal stability (i.e. $\omega^2 = 0$),
\begin{equation}\label{eq:resonance}
	\vrm{k}^{m,n}_{yz}\dotv\vrm{B}(x^{m,n}_{\rm r}) = 0 \;,
\end{equation}
which, by \eqref{eq:Fdef}, can also be written $F_{m,n}(x^{m,n}_{\rm r}) = 0$.

It is well known that, for given $m$ and $n$, such an equilibrium is unstable to an $x,y$ translational symmetry-breaking ``helical bifurcation'' when $|\mu|$ exceeds the first Beltrami eigenvalue $\mu_1^{m,n}$\footnote{Not to be confused with the linear correction to $\mu$ derived in Sec.~\ref{sec:slabeqpert}, which does not apply in the present case with $\vrm{k}_{yz} \neq 0$.} (see e.g. the review by Taylor \cite{Taylor_86}). In slab geometry this corresponds to the lowest value of $|\mu|$ for which a standing wave obeying the boundary conditions $b_x = 0$ [see \eqref{eq:bcs}] at both boundaries $x=0$ and $x = x_{\rm w}$ occurs. That is, from \eqref{eq:bstandingwave},
$k_x^{\mu}x_{\rm w} = \pi$. From \eqref{eq:kxmu} this implies \cite{SuppCitePPCF} 
\begin{equation}\label{eq:BeltramiEigval}
\begin{split}
	\mu_1^{m,n} &= \frac{\pi}{x_{\rm w}}\left[1 + \left(\frac{|\vrm{k}^{m,n}_{xy}|x_{\rm w}}{\pi}\right)^2\right]^{1/2} \;.
\end{split}
\end{equation}

The helical bifurcation can also be identified as due to the tearing mode instability \cite{Taylor_86}. In this section we see how the tearing instability manifests itself in dynamical MRxMHD.

\begin{figure}[htbp]
   \centering
		\includegraphics[width = 0.7\textwidth]{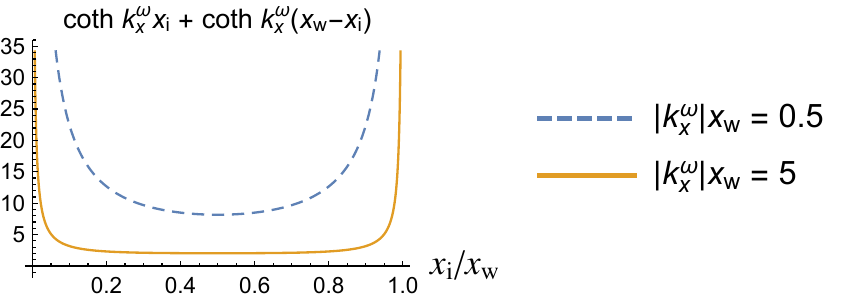} 
\caption{Illustrating the positivity of the $x_{\rm i}$-dependent factor in the ``normalization factor'' $\mathcal{K}$ when $k^\omega_x$ is on the imaginary axis (e.g. in the case of instability, $\omega^2 < 0$).}
\label{fig:tearingcoth}
\end{figure}

In evaluating $\mathcal{K}$ and $\mathcal{W}$ for this equilibrium we can drop ${}^\sglt$ on everything except $a_{\rm i}^\slt = x_{\rm i}$ and $a_{\rm i}^\sgt = x_{\rm w} - x_{\rm i}$. From \eqref{eq:surfwKdefIm}, taking $a^0 = x_{\rm w}$, we find
\begin{equation}
\begin{split}\label{eq:surfwKbif}
	\mathcal{K}(\lambda,x_{\rm i})
	&= \frac{\coth |k_x^{\omega}| x_{\rm i} + \coth |k_x^{\omega}| (x_{\rm w} - x_{\rm i})}{|k_x^{\omega}|x_{\rm w}}
 \;,
\end{split}
\end{equation}
As illustrated in Fig.~\ref{fig:tearingcoth}, $\mathcal{K}$ is positive definite in the case relevant to stability studies, $k^{\sglt\omega}_x = \ijot |k^{\sglt\omega}_x|$. 
From \eqref{eq:kmn} and \eqref{eq:surfwWdef}, 
\begin{equation}\label{eq:surfwWbif}
	\mathcal{W}(\mu,x_{\rm i}) =
	[F^{m,n}(x_{\rm i})]^2\left(\cot k_x^{\mu} a_{\rm i}^\slt
	+ \cot k_x^{\mu} a_{\rm i}^\sgt\right)k_x^{\mu}x_{\rm w} \;.
\end{equation}

\begin{figure}[htbp]
   \centering
		\includegraphics[width = 0.7\textwidth]{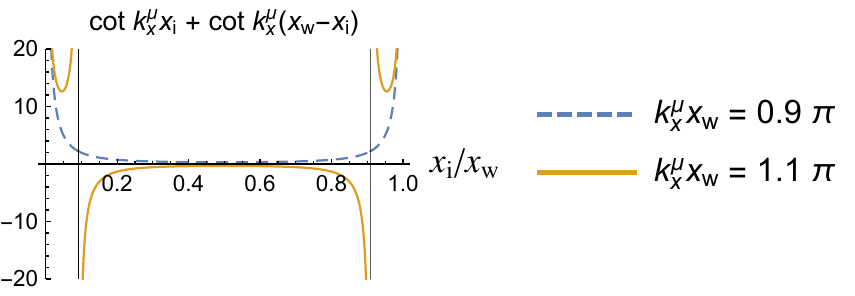} 
\caption{Illustrating how, when $k_x^{\mu}x_{\rm w} > \pi$, there is a range of interface positions $\pi/k_x^{\mu} < x_{\rm i} < x_{\rm w} - \pi/k_x^{\mu}$ over which $\mathcal{W}$ is negative.}
\label{fig:tearingcot}
\end{figure}

From \eqref{eq:surfeigvalue}, $\omega^2$ cannot be negative unless $\mathcal{W}$ is also negative. When we are seeking unstable, or close-to-unstable, modes we can thus restrict attention to the case of \emph{real} $k_x^{\mu}$, as it is easy to show from \eqref{eq:surfwWbif} using the identity in the previous section, 
that, in the case of imaginary $k_x^{\mu}$, $\mathcal{W}(x_{\rm i})$ is positive definite. From \eqref{eq:kxmu} real $k_x^{\mu}$ implies $|\mu| > |\vrm{k}_{yz}|$, but this is not sufficient to make $\mathcal{W}$ negative --- $x_{\rm i}$ must lie within a range of values for which $\cot k_x^{\mu} a_{\rm i}^\slt + \cot k_x^{\mu} a_{\rm i}^\sgt$ is negative, the existence of which requires $|\mu| > \mu_1^{m,n}$ as illustrated in Fig.~\ref{fig:tearingcot}. 

Thus we have shown that our MRxMHD stability analysis captures the onset point of the Taylor bifurcation. Beyond the bifurcation point it gives collisionless tearing mode growth rates that can be regarded as upper bounds for tearing instabilities assisted by the mesoscopic reconnection mechanisms implicit in Taylor relaxation, such as ``chaos-induced resistivity'' \cite{Numata_Yoshida_02}. It is to be noted that placing the interface too close to either wall suppresses the instability and that the growth rate goes to infinity at the edges of the unstable region (where, though outside the scope of the present study, the nonlinearly saturated amplitude presumably goes to zero).

\begin{figure}[htbp]
   \centering
		\includegraphics[width = 0.7\textwidth]{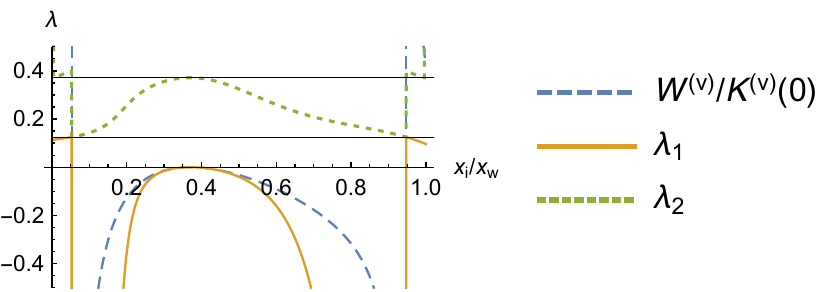} 
\caption{Scan of $\lambda$ \emph{vs.} $x_{\rm i}$ for the RFP-like equilibrium depicted in Fig.~\ref{fig:qprofRFP}, with $\beta = 0.05$ plasma between and $x = x_{\rm i}$ and $x_{\rm w}$. The horizonal guidelines are as in Fig.~\ref{fig:lambdamuscan}.}
\label{fig:lambdaRFP}
\end{figure}

We confirm the predictions of this analysis in Fig.~\ref{fig:lambdaRFP} using the same equilibrium as in Fig.~\ref{fig:qprofRFP}, in the case of plasma filling both regions between the rigid boundaries $x = 0$ and $x_{\rm w}$. As the Beltrami eigenvalue $\mu_1^{-1,5} = 3.86/x_{\rm w}$ for this case is less than the equilibrium value $\mu^\slt = 4/x_{\rm w}$ we expect the plasma to be unstable to the $-1,5$ tearing mode. The lowest eigenvalue $\lambda_1$ shown in Fig.~\ref{fig:lambdaRFP} is indeed negative over a range of interface positions within the plasma. Also shown is the stable second eigenvalue $\lambda_2$ as well as the prediction of the incompressible approximation \eqref{eq:incompressible}, which is seen to be accurate only over a very narrow interval of $\lambda$ for the $5\%$ $\beta$ value used.

Also note that the location $x = x^{m,n}_{\rm r}$ of the mode rational surface does not enter into the instability threshold condition $|\mu| = \mu_1^{m,n}$, but beyond threshold it does affect the dependence of growth rate on interface location $x_{\rm i}$. In fact, the growth rate clearly vanishes at $x_{\rm i} = x^{m,n}_{\rm r}$, consistently with the picture that an MRxMHD interface is a thin layer of ideal plasma, which does not allow reconnection \cite{Yoshida_Dewar_12}. This is apparent in Fig.~\ref{fig:lambdaRFP}  where the growth rate vanishes at $x = x^{-1,5}_{\rm r} = 0.3694 x_{\rm w}$.

\begin{figure}[htbp]
   \centering
		\includegraphics[width = 0.7\textwidth]{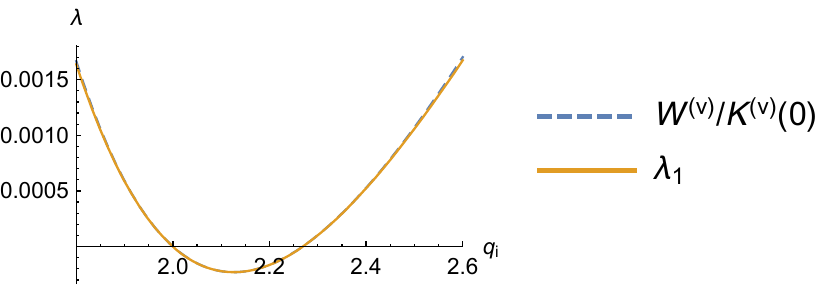} 
\caption{Scan of $\lambda$ \emph{vs.} $q_{\rm i}$ for a family of tokamak-like equilibria iincluding the that depicted in Fig.~\ref{fig:qprofTok}, with $\beta = 0.01$ plasma between $x =0$ and $x_{\rm i}$ and vacuum between and $x = x_{\rm i}$ and $x_{\rm w}$. Parameters are the same as in Fig.~\ref{fig:qprofTok} except for $\mu^\slt$, being a function of $q_{\rm i}$.}
\label{fig:lambdaTok}
\end{figure}

In the tokamak-like case shown in Fig.~\ref{fig:qprofTok}, \eqref{eq:BeltramiEigval} gives $\mu_1^{2,1} = 5.1/x_{\rm w}$, which is far greater than $|\mu^\slt| = 0.16/x_{\rm w}$. In the case where plasma extends continuously to the wall, this tokamak-like equilibrium is therefore stable against the $2,1$ tearing mode (and indeed any mode, as $\mu^\slt < \pi/x_{\rm w}$).

However Fig.~\ref{fig:lambdaTok}  shows that, in the case of vacuum between $x = x_{\rm i}$ and $x_{\rm w}$, there is a narrow range of instability of a $2,1$ mode between $q_{\rm i} = 2$ and $q_{\rm i} \approx 2.23$. As the free-boundary plasma slab is known to be stable against ideal kink modes \cite{Goedbloed_Dagazian_71}, this must be a kink-tearing mode, i.e. a tearing mode made unstable by the proximity of the resonant surface $x = x_{\rm r}^{2,1}$ to the plasma-vacuum interface, $x_{\rm r}^{2,1}$ starting at at $x_{\rm i}$ when $q_{\rm i} = 2$ and moving inward as $q_{\rm i}$ increases. Figure~\ref{fig:lambdaTok} also shows that this instability is so weak that the incompressible approximation gives excellent agreement with the exact solution even with the low $1\%$ $\beta$ used to make the plot. 

\section{Conclusion}\label{sec:Concl}

In order to make clear the physics implications of the new dynamical MRxMHD formulation we have used as elementary an approach as possible, in particular deriving the eigenvalue problem from first principles by linearizing the raw Euler--Lagrange equations arising from the action principle. To develop further insights it remains to derive a quadratic Lagrangian variational principle in terms of the fluid displacement $\bm \xi$, similar to that used in our previous 
energy-principle-based 
MRxMHD formulation \cite{Hole_Hudson_Dewar_07,Mills_Hole_Dewar_09, Hole_Mills_Hudson_Dewar_09}. 

A useful result of the analysis has been to establish that MRxMHD leads to the same stability boundaries (marginal stability points) as our previous formulation by deriving this previous formulation as the incompressible approximation to the new theory, valid near marginal stability.

We have also verified that the instability threshold, derived from linearized MRxMHD in a plasma confined between rigid boundaries, agrees with the onset of the Taylor bifurcation derived from Beltrami equilibrium theory. Given the simplicity of MRxMHD it should also be feasible, and instructive, to calculate nonlinearly saturated amplitudes of linearly unstable modes by expanding the perturbed energy up to quartic order in amplitude, assuming the bifurcation saturates when a point of minimum energy is reached. 
This would be much more physically useful than the rather unphysical growth rates derived in this paper (which, e.g., can become infinite!). 

The single-interface slab model used in the present paper is clearly inadequate for understanding mode structure in a plasma with realistic pressure and current profiles. To find more physical eigenfunctions and eigenvalues we will need to use a realistic geometry and add more interfaces so as to partition the plasma into \emph{multiple} relaxed regions (as already done for MRxMHD equilibria \cite{Hudson_etal_12b}). 

A feature of the dynamical formulation is that it allows treatment of background flow in a natural way, providing a further class of instability mechanisms to explore. After a few more scoping studies in simple geometries we expect MRxMD to provide the basis for an efficient, fully three-dimensional stability and stable-spectrum code.

\section*{Acknowledgments}
One of us (LHT) acknowledges support from the Australian Institute of Physics for a Research Student Travel Grant to present some of this work. The calculations and plots were performed using Mathematica \cite{Mathematica10}.

\section*{References}

\bibliography{MRxMHDrefs}

\end{document}